\documentclass{elsart}
\usepackage{epsfig}
\begin{document}


\newcommand{\bea}{\begin{eqnarray}}
\newcommand{\eea}{\end{eqnarray}}
\newcommand{\be}{\begin{equation}}
\newcommand{\ee}{\end{equation}}
\newcommand{\nn}{\nonumber}
\newcommand{\nnnl}{\nonumber\\}

\newcommand{\ggpp}{\gamma\gamma \rightarrow \pi ^+\pi^-}
\newcommand{\ggppp}{\gamma\gamma \rightarrow \pi ^1\pi^1}

\newcommand{\sss}{\scriptscriptstyle}

\newcommand{\chpt}{ChPT\,\,}
\newcommand{\la}{\;\langle\;}
\newcommand{\ra}{\;\rangle\;}


\renewcommand{\theequation}{\arabic{section}.\arabic{equation}}


\begin{frontmatter}

\title{Revisiting \boldmath{$\gamma\gamma\to\pi^+\pi^-$} 
at low energies}

\author[Bern]{J. Gasser},
\author[Dubna]{M.A. Ivanov},
\author[Helsinki]{M.E. Sainio}

\address[Bern]{Institute for Theoretical Physics, University of Bern,
Sidlerstrasse 5, \\
CH-3012 Bern, Switzerland}
\address[Dubna]{Laboratory of Theoretical Physics,
Joint Institute for Nuclear Research, \\
141980 Dubna (Moscow region), Russia}
\address[Helsinki]{Helsinki Institute of Physics, P.O. Box 64, 00014 
University of Helsinki, Finland and
Department of Physical Sciences, University of Helsinki, Finland}

\begin{abstract}
We complete the recalculation of the available  two-loop expressions 
for the reaction  $\gamma\gamma\to\pi\pi$  in the framework of  
chiral perturbation theory.
Here, we present the results for charged pions.
The cross section and the values of the dipole polarizabilities 
agree very well with the earlier calculation, provided the same set of 
low-energy constants (LECs) is used.
With updated values for the LECs at order~$p^4$,
we find for the dipole polarizabilities   $(\alpha_1-\beta_1)_{\pi^\pm}
= (5.7\pm 1.0)\times 10^{-4}\,{\rm fm}^3$, which is in conflict
 with the experimental result recently reported  by the MAMI Collaboration.

\vspace{0.5cm}
\end{abstract}

\end{frontmatter}

\setcounter{equation}{0}

\section{Introduction}

We evaluate the amplitude for $\gamma\gamma\rightarrow\pi^+\pi^-$ 
in the framework of chiral perturbation theory (\chpt) 
\cite{Weinberg:1978kz,GLann,GLnpb} at two-loop order,
and compare the result with the only previous calculation
performed at this accuracy  \cite{Burgi}. We employ 
the calculational techniques outlined in our previous
work \cite{GIS}, which allow us to provide a rather compact 
 and easy to use integral representation
for the full amplitude. We find that the cross section for the reaction 
$\gamma\gamma\rightarrow\pi^+\pi^-$ and the dipole 
polarizabilities agree very well with the results reported in 
Ref.~\cite{Burgi}, provided that the same set of LECs is used.
 With updated LECs at order $p^4$
 \cite{CGLpipi,BijnensTalavera}, we find 
  for the dipole polarizabilities the value
\bea
\label{eq:polarChPT} 
(\alpha_1-\beta_1)_{\pi^\pm}= (5.7\pm 1.0)\times 10^{-4}\,{\rm fm}^3\,.
\eea
The MAMI Collaboration \cite{MAMI} has recently 
reported the experimental result
\bea
\label{eq:polarMAMI}
(\alpha_1-\beta_1)_{\pi^\pm}= (11.6\pm 1.5_{\rm stat}
\pm 3.0_{\rm syst}\pm 0.5_{\rm mod})\times 10^{-4}\,{\rm fm}^3\,.
\eea
The index ``mod'' denotes
the uncertainty generated by the theoretical models used to analyse
the data. The ChPT calculation is in conflict with this
 prediction, see 
 also \cite{schererreview} for a recent discussion.

There are good reasons to believe that the chiral prediction 
Eq.~(\ref{eq:polarChPT}) is rather stable against contributions
 from still higher orders in the chiral expansion. 
The main point is that the Born-term
subtracted amplitude - in terms of which the polarizabilities are defined
 - is regular at the Compton threshold.
Therefore, the influence of chiral logarithms is
suppressed as compared to the amplitude at, say, the threshold for
$\pi\pi$ production, where the amplitude generates a branch point.
Further, we have analysed the potential influence of resonance exchange
and could not identify any significant contribution from nearby resonances
 to the combination $(\alpha_1-\beta_1)_{\pi^\pm}$.
 We conclude that the low-energy constants at order~$p^6$
are expected to play a negligible effect here.
In view of these observations, the discrepancy between the chiral
prediction and the recent MAMI data cannot be explained.

The article is organized as follows. In Section 2 we spell out 
the kinematics of the process $\gamma\gamma\rightarrow\pi^+\pi^-$.
To make the article selfcontained, we summarize in Section 3 the necessary
ingredients of the effective  Lagrangian framework. Here, we also fix the
LECs to be used in numerical calculations. In Section 4, 
we display the  Feynman diagrams and discuss shortly their evaluation. 
Section 5 contains a concise representation of the two Lorentz 
invariant amplitudes that describe the scattering matrix element.
Section 6 contains explicit expressions for the dipole and 
quadrupole polarizabilities valid at next-to-next-to-leading order 
in the chiral expansion, together with a detailed numerical analysis and 
a comparison with the recent MAMI data \cite{MAMI}, 
and with  an evaluation from data on 
$\gamma\gamma\rightarrow\pi^+\pi^-$~ \cite{FK05}.
The summary and an outlook are given in Section 7.
A detailed comparison with the earlier calculation \cite{Burgi} 
is provided in Appendix~\ref{app:comparison}, and several technical aspects 
of the calculation are relegated to additional
 Appendices~\ref{app:loopfunc}-\ref{app:polynomials}.

\setcounter{equation}{0}

\section{Kinematics\label{nk}}

The amplitude describing the process $\ggpp$ may be extracted from the matrix
element
\bea
\hspace*{-0.7cm}
\la \pi^+(p_1)\pi^-(p_2)
\;{\rm out}\;|\;\gamma(q_1)\gamma(q_2) \; {\rm in} \ra 
= i (2\pi)^4\delta^4(P_f-P_i) \;T^{\ggpp},
\label{kin1}
\eea
where
\bea
\hspace*{-0.7cm}
T^{\ggpp} & = & e^2
\epsilon_1^{\mu}\epsilon_2^{\nu}\,
W_{\mu\nu}^{\ggpp} \; , 
\nn \\ 
\hspace*{-0.7cm}
W_{\mu\nu}^{\ggpp} & = & 
i \int \! dx \, e^{-i(q_1 x+q_2 y)} 
\la\pi^+(p_1)\pi^-(p_2) \;{\rm out} \;| \;T  j_{\mu}(x) j_{\nu} (y) \;
| \; 0\ra.
\label{kin2}
\eea
Here $j^{\mu}$ denotes the electromagnetic current and $\alpha=e^2/4\pi \simeq
1/137$ is the  electromagnetic coupling.  It is convenient to change the pion
coordinates according to $(\pi^{\pm},\pi^0) \rightarrow (\pi^1,\pi^2,\pi^3)$ and
instead of  $\pi^+\pi^-$--production, we consider in the following the process
$\gamma\gamma\rightarrow \pi^1\pi^1$, with
\bea
W_{\mu\nu}^{\ggpp} = -W_{\mu\nu}^{\ggppp} \doteq -V_{\mu\nu} \; ,
\label{kin3}
\eea
where the relative minus sign stems from the Condon--Shortly phase convention.
[We use the same sign convention as Ref.~\cite{Burgi}.]
The decomposition of the correlator
$V_{\mu \nu}$ into Lorentz invariant amplitudes reads
\begin{eqnarray}
V_{\mu \nu} &=&  A(s,t,u) T_{1\, \mu \nu} + B(s,t,u) T_{2\,\mu \nu} 
               + C(s,t,u) T_{3\, \mu\nu}  + D(s,t,u) T_{4\, \mu \nu} \,,
\label{eq:amplitude}\nonumber\\
&&\nonumber\\
T_{1\, \mu \nu} &=& \frac{1}{2}\,s\, g_{\mu \nu} - q_{1\nu} q_{2\mu} \,,
\nonumber\\
&&\nonumber\\
T_{2\, \mu \nu} &=& 2\,s\,\Delta_\mu\Delta_\nu - \nu^2\,g_{\mu\nu}
                   -2\,\nu\, (q_{1\,\nu}\Delta_\mu - q_{2\,\mu}
\Delta_\nu) \,,
\nonumber\\
&&\nonumber\\
T_{3\, \mu \nu} &=& q_{1\,\mu} q_{2\,\nu} \,, 
\nonumber\\
&&\nonumber\\
T_{4\, \mu \nu}  &=&  s\,(q_{1\, \mu} \Delta_\nu - q_{2\,\nu} \Delta_\mu)
         - \nu\,(q_{1\, \mu} q_{1\, \nu} + q_{2\,\mu} q_{2\,\nu}) \,,
\nnnl
&&\nnnl
\Delta_\mu &=& (p_1 -p_2)_\mu\, ,
\end{eqnarray}
where
\bea
s &=& (q_1 + q_2)^2,\;\;\; t = (p_1 -q_1)^2, \;\;\; u = (p_2 - q_1)^2, \;\;\;
\nu = t-u \,
\eea
are the standard Mandelstam variables. The tensor $V_{\mu\nu}$  
satisfies the Ward identities
\begin{equation}
\label{eq:ward}
q^\mu_1\, V_{\mu \nu} = q^\nu_2\, V_{\mu \nu} = 0.
\end{equation}
The amplitudes $A$ and $B$  are analytic functions of the variables 
$s,t$ and $u$, symmetric under crossing $(t,u)\rightarrow (u,t)$. 
The amplitudes  $C$ and $D$ do not contribute to the process considered here,
because $\epsilon_i\cdot q_i=0$.

It is useful to introduce in addition the helicity amplitudes
\begin{eqnarray}
\label{eq:hel}
{\bar H}_{++} &=& A + 2\,(4\,M^2_\pi - s)\, B \,,
\qquad
{\bar H}_{+-} =  \frac{8\,(M_\pi^4- t\,u)}{s}\, B.
\end{eqnarray}
The helicity components ${\bar H}_{++}$ and ${\bar H}_{+-}$ correspond to
photon helicity differences $\lambda = 0,2$, respectively.
 With our normalization of 
states $\langle\mathbf{p_1} | \mathbf{p_2}\rangle
 = 2\,(2 \pi)^3\, p_1^0\, \delta^{(3)} (\mathbf{p_1} - \mathbf{p_2})$, the
differential cross section for unpolarized photons in the
centre-of-mass system is
\begin{eqnarray}
\label{eq:cross}
{\frac{d \sigma}{d \Omega}}^{\gamma \gamma \rightarrow \pi^+ \pi^-} 
&=& \frac{\alpha^2\,s}{32} \beta(s)\, H(s,t)\,,
\hspace{1cm}
H(s,t) = | {\bar H}_{++}|^2 + | {\bar H}_{+-}|^2 \,,
 \end{eqnarray}
with $\beta(s)=\sqrt{1-4\, M^2_\pi/s}$.
 The relation between the helicity 
amplitudes $M_{+\pm}$ in Ref. \cite{FK05} and 
the amplitudes used here is
\bea
M_{++}(s,t)=2\pi\alpha {\bar H}_{++}(s,t)\,,\qquad
M_{+-}(s,t)=16\pi\alpha B(s,t)\,.
\eea

In the centre-of-mass system, $\vec q_1+\vec q_2 =0$, one has
$\vec q_1\cdot \vec p_1=|\vec q_1||\vec p_1|\cos\theta$, where
$\theta$ is the scattering angle. Then the Mandelstam variables
are given by

\bea
s=4\,|\vec q|^2, \qquad t=M^2_\pi-(s/2)\left(1-\beta(s)\cos\theta\right).
\eea
For comparison with experimental data, it is convenient to present
also the total cross section for the case having $|\cos\theta|$ less
than some fixed value $Z$,

\bea
\sigma(s;|\cos\theta|<Z) = \frac{\alpha^2\pi}{8}
\int\limits_{t_-}^{t_+} dt H(s,t) 
\eea
with $t_{\pm} = M_{\pi}^2- (s/2)\; (1 \mp \beta(s) Z)$.

\setcounter{equation}{0}
\section{The effective Lagrangian and its low-energy constants} 
\label{sec:effective}
The  effective Lagrangian consists of a string of terms. 
Here, we consider QCD with two flavours, in the
isospin symmetry limit $m_u=m_d=\hat m$. At 
next-to-next-to-leading order (NNLO), one has \cite{GLann}
\bea
{\mathcal L}_{\rm eff}={\mathcal L}_2+{\mathcal L}_4 +{\mathcal L}_6\,.
\eea
The subscripts refer to the chiral order. 
The expression for ${\mathcal L}_2$ is
\bea
\label{eq:l2}
{\mathcal L}_{\, 2} &=&\frac{F^2}{4}\langle D_\mu U\,D^\mu U^\dagger
        +M^2(U + U^\dagger)\rangle\, ,\nnnl
D_\mu U &=& \partial_\mu U -i(QU-UQ)A_\mu\, , \, \
Q=\frac{e}{2}{\rm diag}(1,-1)\, ,
\eea
where $e$ is the electric charge, and $A_\mu$ denotes the electromagnetic 
field.
The quantity $F$ denotes the pion decay constant 
in the chiral limit,  and $M^2$ is the leading term in the quark mass
expansion of the pion (mass)$^2$, $M_\pi^2=M^2(1+O(\hat m))$.
Further, the brackets $\langle\ldots\rangle$ denote a trace in flavour space.
In Eq. (\ref{eq:l2}), we have retained only the terms relevant for 
the present application,
i.e., we have dropped additional external fields.
We choose the unitary $2\times 2$ matrix $U$ in the form
\bea
U &=& \sigma + i\, \pi/F\,, 
\hspace{.3cm} \sigma^2 + \frac{\pi^2}{F^2} = {\mathbf 1}_{2\times 2}\,,
\hspace{.3cm}\pi=\left( \mbox{$\begin{array}{cc} \pi^0 & \sqrt{2}\, \pi^+
         \\ \sqrt{2}\, \pi^- &- \pi^0 \end{array}$} \right)  \; .
\eea
The  Lagrangian at NLO  has the structure~\cite{GLann}
\bea
{\mathcal L}_4=\sum_{i=1}^{7} l_iK_i+\sum_{i=1}^{3}h_i\bar K_i=\frac{l_1}{4}\langle
D_\mu U\,D^\mu U^\dagger\rangle^2
+\cdots\, ,
\eea
where  $l_i,h_i$ denote low-energy couplings, 
not fixed by chiral symmetry.  
At NNLO, one has \cite{BCE1,BCE2,SchererFearing}
\bea
{\mathcal L}_6=\sum_{i=1}^{57} c_i P_i\,.
\eea
For the explicit expressions of the polynomials $K_i,\bar K_i$ and $P_i$, 
we refer the reader to
Refs.~\cite{GLann,BCE1,BCE2,SchererFearing}. The vertices relevant for
$\gamma\gamma\rightarrow \pi^+\pi^-$ involve  $l_1, \ldots, l_6$ 
{}from ${\mathcal L}_4$ and several $c_i$'s 
{}from ${\mathcal L}_6$, see below.

The couplings $l_i$ and $c_i$ absorb the divergences at order~$p^4$ and 
$p^6$, respectively,
\bea\label{eq:lici}
l_i &=& (\mu\,c)^{d-4}
\left\{l_i^r(\mu,d) + \gamma_i\,\Lambda\right\}\,, 
\nn\\[2mm]
c_i&=&\frac{(\mu\,c)^{2(d-4)}}{F^2}
\left\{
c_i^r(\mu,d) - \gamma_i^{(2)}\,\Lambda^2
       -(\gamma_i^{(1)}+\gamma_i^{(L)}(\mu,d))\,\Lambda\right\}\,, 
\nn\\[2mm]
\Lambda &=& \frac{1}{16\,\pi^2 (d-4)}\,,
\, \ln c = -\frac{1}{2}\left\{\ln 4\pi +\Gamma'(1)+1\right\}\,.
\eea
The physical couplings are $l_i^r(\mu,4)$ and $c_i^r(\mu,4)$, denoted by
$l_i^r,c_i^r$ in the following.
The coefficients $\gamma_i$ are given in \cite{GLann}, 
and $\gamma_i^{(1,2,L)}$ are tabulated in \cite{BCE2}.
 In order to compare the present calculation with the result of
\cite{Burgi}, we shall use the scale independent quantities 
$\bar l_i$ introduced in \cite{GLann}, 
\begin{eqnarray} 
\label{RGE}
l_i^r &=& \frac{\gamma_i}{32\pi^2}\,({\bar l}_i + l)\,,
\end{eqnarray}
where the {\it chiral logarithm} is $l=\ln(M^2_\pi/\mu^2)$. We shall 
use \cite{CGLpipi}
\bea
\label{eq:LECsp4}
\bar l_1=-0.4\pm 0.6\,,\,\,\bar l_2=4.3\pm0.1\,,\,\,
\bar l_3=2.9\pm2.4\,,\,\,
\bar l_4=4.4\pm0.2\,,
\eea
and 
\bea
\label{eq:LECs65}
\bar l_\Delta\doteq \bar l_6-\bar l_5=3.0\pm 0.3
\eea
obtained from radiative pion decay to two loop
accuracy \cite{BijnensTalavera,Geng}.

The constants $c_i^r$ occur in the combinations
 \begin{eqnarray}
a_1^r &=& -4096\pi^4\left( 6\, c_{6}^r+ c_{29}^r -c_{30}^r -3\, c_{34}^r 
          + c_{35}^r + 2\,c_{46}^r - 4\, c_{47}^r + c_{50}^r \right)\,,
\nn\\
&&\nn\\
a_2^r &=& 256\pi^4\left( 
8\,c_{29}^r - 8\,c_{30}^r + c_{31}^r + c_{32}^r- 2\,c_{33}^r 
+4\, c_{44}^r +8\, c_{50}^r -4\, c_{51}^r \right)\,,
\nn \\
&&\nn\\
b^r &=&-128\pi^4
     \left(c_{31}^r + c_{32}^r - 2\,c_{33}^r -4\, c_{44}^r \right)\,.\nonumber
\end{eqnarray}
Their values have been estimated by resonance exchange e.g. in 
Ref.~\cite{Burgi}. We have repeated that analysis. Taking into account
 $\rho, a_1 $ and $b_1$ exchange which contribute with a definite sign, 
we  obtain
\bea\label{eq:LECsp6}
\Big(a_1^r, a_2^r, b^r\Big)=
\Big(-3.2,0.7,0.4\Big)\qquad  [{\mbox{ present work }}]\,.
\eea
Unless stated otherwise, we will use these estimates 
at the scale $\mu=M_\rho$.
 Contributions from scalar and tensor  exchange are of a similar order of
magnitude (see Table~2 in Ref.~\cite{BGS}).
 On the other hand, 
the sign of these contribution is not fixed. 
In Ref.~\cite{BijnensPrades}, a large $N_C$ framework and the ENJL
model were used 
to pin down these constants, with the result
\bea
\label{eq:LECsp6bijnensprades}
\Big(a_1^r, a_2^r, b^r\Big)=
\Big(-8.7,5.9,0.38\Big)\qquad {\mbox{Ref.~\cite{BijnensPrades}}}\,.
\eea
Only $b^r$ agrees in the two approaches. We have checked that scalar and
tensor exchange, taken with the proper sign, generate values for 
$a_{1,2}^r$ that are not in
disagreement with Eq.~(\ref{eq:LECsp6bijnensprades}) - as has been 
foreseen  in the comments made in Ref.~\cite{BijnensPrades} 
concerning these
two approaches. It would be very useful to
recalculate  these couplings, by minimizing the amount of information used
which  goes beyond what is known from QCD, e.g., along the lines outlined in 
\cite{LECs_asymptotic}.
Finally, we note that in Ref.~\cite{KnechtMoussallamStern}, $c_{34}^r$
has been determined from a chiral sum rule. 

We  now shortly discuss the uncertainties that we shall attach
in the following to these couplings.
 In the case of  $b^r$, we shall use
\bea\label{eq:br}
b^r=0.4\pm 0.4\,.
\eea
 As far as the polarizabilities are concerned, 
$(\alpha_1-\beta_1)_{\pi^\pm}$ is independent of $a_2^r$ and determined 
precisely by the chiral expansion to two loops,
once $a_1^r$ is fixed. We will then simply display this quantity as
 a function of $a_1^r$ - the result turns out to be 
 rather independent of its exact
 value, see Subsection~\ref{subsec:uncertainties} 
for a detailed discussion, and the  uncertainty to be attached 
to it does not, therefore, matter here. On the other hand, 
those polarizabilities which depend on
 $a_2^r$ cannot be determined precisely from a 
calculation to two loops for reasons explained in 
Subsection~\ref{subsec:uncertainties} - we
 do not, therefore, worry here about the precise value and uncertainty 
 for $a_2^r$.

To complete this discussion, we note  that  we 
shall  use $F_\pi=92.4$ MeV~\cite{Holsteinfpi} 
(see \cite{Descotesfpi} for a recent update of this value), 
and $M_\pi = 139.57$ MeV in numerical calculations.

\setcounter{equation}{0}

\section{Evaluation of the diagrams}
\label{sec:diagrams}

\begin{figure}[!ht]
\begin{center}
\epsfig{figure=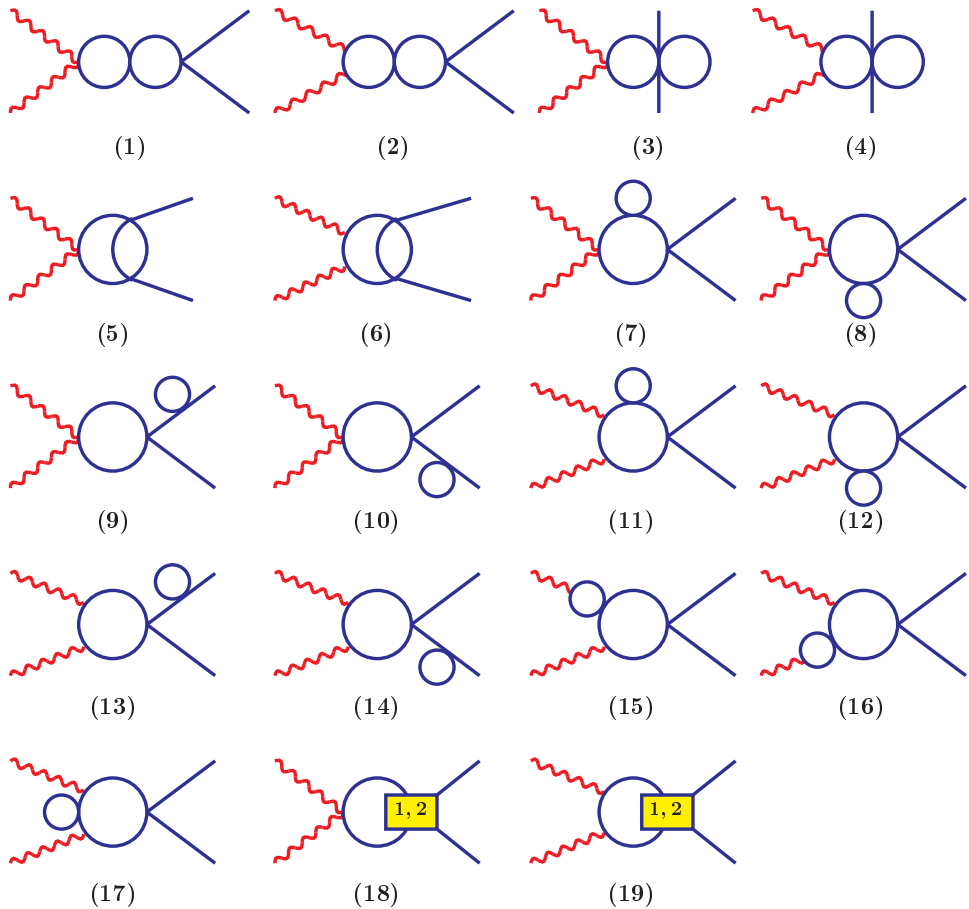,width=13.5cm}
\end{center}
\caption{\small
A set of two-loop diagrams generated by ${\mathcal L}_2$
and one-loop diagrams generated by ${{\mathcal L}_4.}$} 
\label{fig:two-loop}
\end{figure}

\begin{figure}[!ht]
\begin{center}
\epsfig{figure=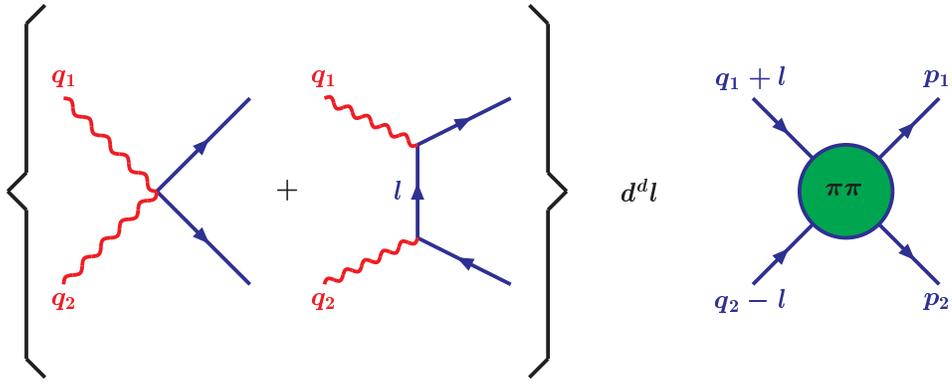,width=13.5cm} 
\end{center}
\caption{\small
Construction scheme for the diagrams in Fig.~\ref{fig:two-loop}.} 
\label{fig:scheme}
\end{figure}

\begin{figure}[!ht]
\begin{center}
\epsfig{figure=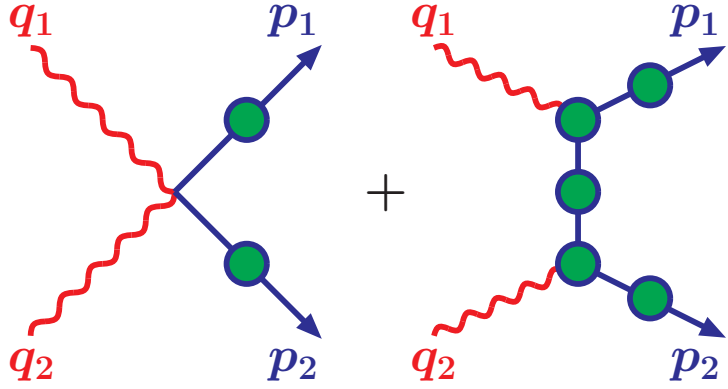,width=10cm} 
\end{center}
\caption{\small A class of one-particle  reducible diagrams. The filled in circles summarize
self-energy and vertex corrections.} 
\label{fig:reducible}
\end{figure}

\begin{figure}[!ht]
\begin{center}
\epsfig{figure=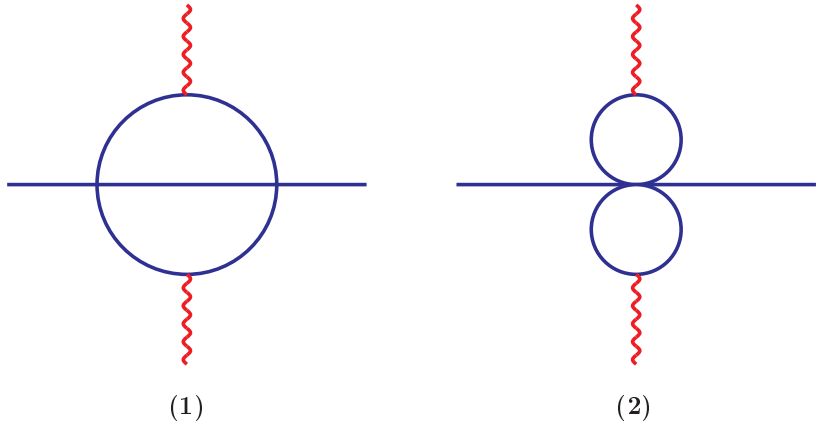,height=6cm} 
\end{center}
\caption{\small Acnode and butterfly diagrams.}
\label{fig:acnode}
\end{figure}

The lowest-order contributions to the scattering amplitude 
are described by  tree- and one-loop  diagrams. 
These contributions were calculated in \cite{BijnensCornet}.
The two-loop diagrams are displayed in the 
Figs.~\ref{fig:two-loop}, \ref{fig:reducible} and \ref{fig:acnode}.
The  two-loop diagrams in Fig.~\ref{fig:two-loop} may be generated 
according to the scheme indicated in 
Fig.~\ref{fig:scheme}, where the filled in blob denotes
the $d$-dimensional elastic $\pi\pi$-scattering amplitude   
at one-loop accuracy, with two pions off-shell. 

The diagrams shown in Fig.~\ref{fig:reducible} 
may be reduced to tree-diagrams by using Ward
identities \cite{Burgithesis}. They sum up to the expression
\bea
&&
2\, Z_\pi\, g^{\mu\nu} 
- \left\{(2 p_1-q_1)^\mu (2 p_2-q_2)^\nu 
         \left[\frac{1}{M^2_\pi-t}-Z_\pi R(t)\right] 
+ {\rm crossed } \right\},
\eea
where $Z_\pi$ is the pion renormalization constant. The function
$R(t)$ starts at order $1/F^4_\pi$ and can be obtained from
the full pion propagator \cite{Burgithesis}.

\begin{figure}[!ht]
\begin{center}
\epsfig{figure=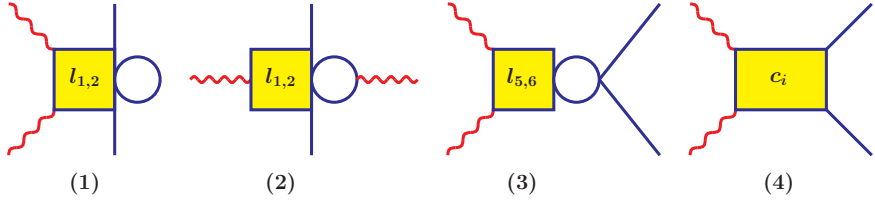,width=12cm}  
\end{center}
\caption{\small The remaining diagrams at order~$p^6$: 
one-loop graphs generated by ${\mathcal L}_4$, and counterterm 
contributions  from ${\mathcal L}_6$.} 
\label{fig:lecs}
\end{figure}

Two further diagrams are displayed in Fig.~\ref{fig:acnode}.
The first one - called  ``acnode'' in the literature - may again be evaluated
by use of a dispersion relation, see \cite{GIS}.
The second one is trivial to evaluate, because it is a product of one-loop
diagrams. 
The remaining diagrams at order~$p^6$ are shown 
in Fig.~\ref{fig:lecs}.

The evaluation of the diagrams was done in the manner
described in \cite{GIS,GS} and invoking  FORM \cite{Vermaseren}.
In particular, we have verified that the
counterterms from the Lagrangian ${\mathcal L}_6$ \cite{BCE2} 
remove all ultraviolet divergences, which is a very non-trivial 
check on our calculation. Furthermore, 
we have checked that the (ultra-violet finite) amplitude so obtained 
is scale independent.

\vspace*{0.5cm}
\setcounter{equation}{0}

\section{The two-loop amplitudes}
\label{sec:2-loop}
We give the expressions for the amplitudes $A$ and $B$ 
using the same notation as in \cite{Burgi}.
This results in

\bea
\hspace*{-0.5cm}
A = \frac{1}{M_{\pi}^2-t}+\frac{1}{M_{\pi}^2-u}
  +\frac{2}{F_{\pi}^2}\left\{\bar{G}_{\pi}(s)
  +\frac{\bar{l}_{\Delta}}{48 \pi^2} \right\} 
  + U_{\sss{A}} + P_{\sss{A}} + {\mathcal O}(p^4).
\label{2amp3}
\eea
The unitary part $U_{\sss{A}}$ contains $s$, $t$ and $u$-- channel
cuts, and $P_{\sss{A}}$ is a linear polynomial in $s$. Explicitly we
find,
\bea
&&
U_{\sss{A}}  =   \frac{1}{s F_{\pi}^4} \bar{G}(s) \; 
\left[ \; (2 M_{\pi}^4-4 M_{\pi}^2 s + 3 s^2) \bar{J}(s) +
      C(s,\bar{l}_i) \; \right] \;
     +\frac{\bar{l}_{\Delta}}{48 \pi^2 F_{\pi}^4} s \bar{J}(s) \nn\\
&&
 + \frac{(\bar{l}_1-\frac{4}{3})}{288 \pi^2 s F_{\pi}^4}
(s-4 M_{\pi}^2) \left\{\bar{H}(s)+4 \;[\;s\bar{G}(s)
+2 M_{\pi}^2(\stackrel{=}{G}(s)-3\stackrel{=}{J}(s)) \;]\;
d_{00}^2 \right\} \nn \\ 
&& +\;
 \frac{(\bar{l}_2-\frac{5}{6})}{96 \pi^2 s F_{\pi}^4} \;
   (s-4 M_{\pi}^2) \left\{\bar{H}(s)+4 \;[\;s\bar{G}(s)
+2 M_{\pi}^2(\stackrel{=}{G}(s)-3\stackrel{=}{J}(s)) \;]\;
d_{00}^2 \right\} \nn \vspace{.5cm} \\ 
& & + \;\Delta_{\sss{A}}(s,t,u) \; ,
\label{2amp4}
\eea
with
\bea
C(s,\bar{l}_i) & = & \frac{1}{48 \pi^2} 
\left\{\frac{1}{3}\left(\bar{l}_1-\frac{4}{3}\right)(16 s^2-56 M_{\pi}^2 s 
+64 M_{\pi}^4)
\right.\nn\\
&& 
+\left(\bar{l}_2-\frac{5}{6}\right)(8 s^2-24 M_{\pi}^2 s +32 M_{\pi}^4) \nn \\
&& 
\left.
-12 M_{\pi}^4 \bar{l}_3 +12 M_{\pi}^2 s \bar{l}_4-12 M_{\pi}^2 s 
+12 M_{\pi}^4 \right\} \; , \nn \\
d_{00}^2 & = & \frac12(3\cos^2\theta-1) \; .
\label{2amp5}
\eea
The loop functions $\bar{J}$ etc. are displayed in 
Appendix~\ref{app:loopfunc}, and
$\bar{G}_{\pi}(s)$ in Eq.~(\ref{2amp3}) stands for $\bar{G}(s)$ evaluated with
the physical mass. The term proportional to $d_{00}^2$ in
$U_{\sss{A}}$ contributes to $D$--waves only.
For $\Delta_{\sss{A}}$ see below. The polynomial part is
\bea
P_{\sss{A}} & = & \frac{1}{(16\pi^2F_{\pi}^2)^2} \,
\left[\,a_1  M_{\pi}^2+a_2 s\,\right]\, \; , \nn \\ 
a_1 & = & a_1^r
+\frac{1}{9}\left\{4\, l^2+l\,\left( -10\,\bar{l}_1
 +18\,\bar{l}_2-12\,\bar{l}_{\Delta}
 +\frac{337}{6} \right)
\right.
\nnnl
&&
\left.
 -\frac{5}{3}\,\bar{l}_1-5\,\bar{l}_2+12\,\bar{l}_4\bar{l}_{\Delta}+4 \right\} 
\; , \nnnl
 a_2 & = & a_2^r-\frac{1}{9}
  \left\{l^2+l\,\left(\frac{1}{2}\,\bar{l}_1+\frac{3}{2}\,\bar{l}_2
 +3\,\bar{l}_{\Delta}+\frac{127}{24}\right)
\right.
\nnnl
&&
\left.
 -\frac{5}{12}\,\bar{l}_1-\frac{5}{4}\,\bar{l}_2+3\,\bar{l}_{\Delta}
+\frac{21}{2} \right\}.
\label{2amp6}
\eea
The result for $B$ reads
\bea
B = \frac{1}{2s}\left\{\frac{1}{M_{\pi}^2-t}+\frac{1}{M_{\pi}^2-u}\right\} 
+ U_{\sss{B}} + P_{\sss{B}} + {\mathcal O}(p^2) \; ,
\label{2amp8}
\eea
with the unitary part
\bea
U_{\sss{B}} & = & \frac{1}{192 \pi^2 s F_{\pi}^4} 
 \left\{ \frac{1}{3}\left(\bar{l}_1-\frac{4}{3}\right)
+\left(\bar{l}_2-\frac{5}{6}\right) \right\}\bar{H}(s) 
+ \Delta_{\sss{B}}(s,t,u) \; .
\label{2amp9}
\eea
For the polynomial part we find
\bea
P_{\sss{B}} & = & \frac{b}{(16\pi^2 F_{\pi}^2)^2} \; , \nn \\ 
b & = & b^r-\frac{1}{18} \left\{ l^2
+l\, \left( \frac{1}{2}\,\bar{l}_1+\frac{3}{2}\,\bar{l}_2
      -\frac{53}{24}\right)-\frac{1}{12}\,\bar{l}_1
-\frac{1}{4}\bar{l}_2+\frac{7}{2} \right\} \; .
\label{2amp10}
\eea
The integrals $\Delta_{\sss{A,B}}(s,t,u)$ contain contributions from
the two-loop box, vertex and acnode graphs and also from the reducible
diagrams. The explicit expressions for these quantities are given
in Appendices~\ref{app:deltaab} and \ref{app:polynomials}
\footnote{The corresponding Fortran codes are available upon request
from the authors.}.

\begin{figure*}
\begin{center} 
\epsfig{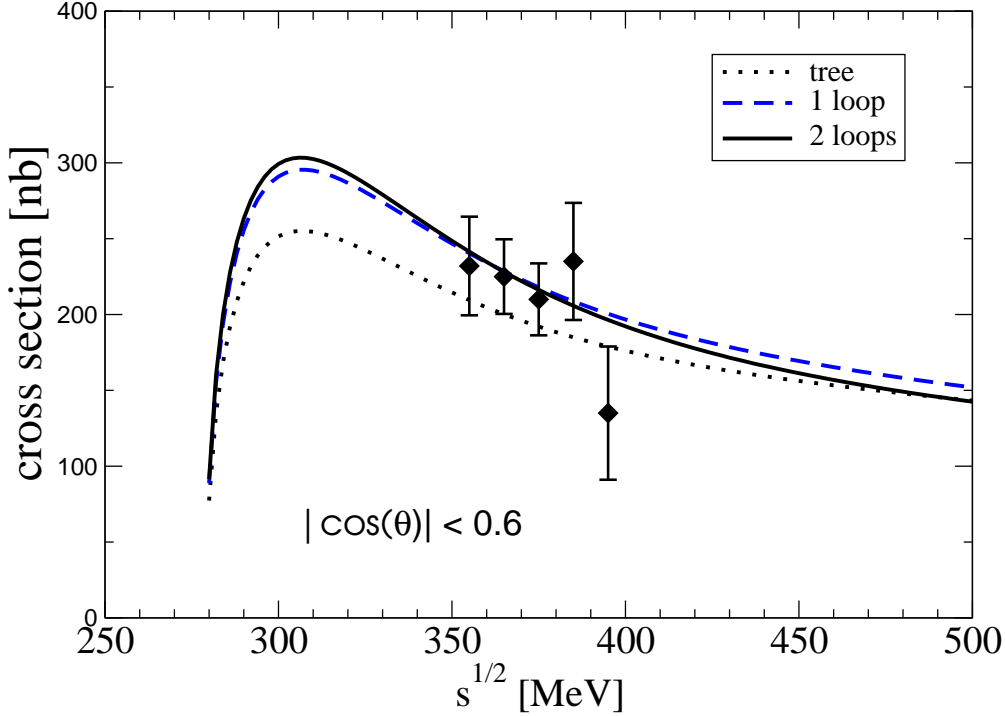} 
\caption{\small The $\gamma\gamma\to\pi^+\pi^-$ cross section
$\sigma(s;\,|\cos\theta| \le Z=0.6)$ as a function of $\sqrt{s}$.
The experimental data are taken from \cite{MARK-II}.        }
\label{fig:cross}
\end{center}
\end{figure*}

As an application of the above, we plot the total cross section 
in Fig.~\ref{fig:cross}, using the LECs from 
Eqs.~(\ref{eq:LECsp4}) - (\ref{eq:LECsp6}). The data 
are taken from \cite{MARK-II}.
It is seen that the two-loop corrections are tiny in this 
kinematical region.

In order not to interrupt the argument, a
detailed comparison of our result with the previous calculation 
performed by Burgi \cite{Burgi} is relegated to 
Appendix~{\ref{app:comparison}}.

\vspace*{0.5cm}

\setcounter{equation}{0}
\section{Pion polarizabilities: dipole and quadrupole}
\label{sec:polar}
The {\it dipole} and {\it quadrupole} 
polarizabilities 
are defined \cite{Guiasu-Radescu,Filkovquadrupole}
through the expansion of the helicity
amplitudes at fixed $t=M_\pi^2$,
\bea\label{eq:defpolarizabilities}
\frac{\alpha}{M_\pi}H_{+\mp}(s,t=M_\pi^2)=(\alpha_1\pm\beta_1)_{\pi^+}
+\frac{s}{12}(\alpha_2\pm\beta_2)_{\pi^+}+{\mathcal O}(s^2)\,,
\eea
where $H_{+\mp}$ denote the helicity amplitudes ${\bar H}_{+\mp}$
with Born-term subtracted.
Because we have at our disposal the helicity amplitudes at two-loop order,
we can work out the polarizabilities to the same accuracy. 
It turns out that all relevant integrals can be performed in closed
form. We discuss the results in the remaining part of this Section.

\subsection{Chiral expansion}
Using the same notation as in \cite{Burgi}, we find for the 
{\it dipole} polarizabilities
\begin{equation}
\label{eq:alphabeta}
(\alpha_1 \pm \beta_1)_{\pi^+} = 
           \frac{\alpha}{16\,\pi^2\,F_\pi^2\,M_\pi}\,
\left\{c_{1\pm} + \frac{M_\pi^2\,d_{1\pm}}{16\,\pi^2\,F_\pi^2} +O(M^4_\pi)
\right\}\,,
\end{equation}
where 
\begin{eqnarray}
c_{1+} &=& 0, \qquad c_{1-}=\frac23\,\bar l_\Delta\, ,
\nnnl
d_{1+} &=& 8\,b^r 
-\frac{4}{9}\left\{ l\,\left( l+\frac{1}{2}\,\bar{l}_1
                   +\frac{3}{2}\,\bar{l}_2\right)
-\frac{53}{24}\,l  + \frac{1}{2}\,\bar{l}_1
+\frac{3}{2}\,\bar{l}_2+\frac{91}{72} +\Delta_+ \right\} \;\; , 
\nnnl
d_{1-} & = & a_1^r +8\,b^r  -\frac{4}{3}
  \left\{l\left(\bar{l}_1-\bar{l}_2+\bar{l}_{\Delta}-\frac{65}{12}\right)
 - \frac{1}{3}\,\bar{l}_1 -\frac{1}{3}\,\bar{l}_2+\frac{1}{4}\,\bar{l}_3
 -\bar{l}_{\Delta}\bar{l}_4
\right. 
\nnnl
&&
\left.
+\frac{187}{108} + \Delta_-\right\}\,,
\label{polchi3}
\eea
with
\bea
\Delta_+ & = & \frac{8105}{576} - \frac{135}{64}\,\pi^2\,, \qquad
\Delta_-   =   \frac{41}{432} - \frac{53}{64}\,\pi^2\,. 
\label{polchi4}
\eea

For the {\it quadrupole} polarizabilities, we obtain
\begin{equation}
\label{alphabeta2}
(\alpha_2 \pm \beta_2)_{\pi^+} = 
           \frac{\alpha}{16\,\pi^2\,F_\pi^2\,M_\pi^3}\,
\left\{c_{2\pm}
 + \frac{M_\pi^2 d_{2\pm} }{16\,\pi^2\,F_\pi^2} +O(M^4_\pi)
\right\}\,,
\end{equation}
with
\bea\label{eq:d2plusminus}
 c_{2+}&=& 0\,,\qquad c_{2-}=2\, ,\nnnl
&&\nn\\
d_{2+}&=&
 - \frac{2062}{27} + \frac{10817}{1440}\,\pi^2 
 + \frac{8}{45}\,\bar l_1 + \frac{8}{15}\,\bar l_2\,,
\nn\\
d_{2-}&=&  12\,a_2^r - 24\,b^r
  -\,l\,(  10 + 4\,\bar l_\Delta )
   - \frac{8}{15}\,\bar l_3   + 4\,\bar l_4 - 4\,\bar l_\Delta 
    - \frac{218}{45}\,\bar l_1    - \frac{238}{45}\,\bar l_2
\nn\\
&&
-\,\frac{56}{45}- \frac{1199}{1920}\,\pi^2\,. 
\eea
We end this subsection by evaluating the polarizabilities, using the above
expressions and the central values for the LECs
in Eqs.~(\ref{eq:LECsp4})-(\ref{eq:LECsp6}).
The results are shown in Table~\ref{tab:polar1}\footnote{Dipole (quadrupole)
  polarizabilities are given in units of $10^{-4}$ fm$^3$
 ($10^{-4}$ fm$^5$).}.
The numbers in brackets correspond to the order~$p^6$ LECs in Eq.~(\ref{eq:LECsp6bijnensprades}).
The uncertainties are discussed in the following subsection.
\begin{table}[ht]
\begin{center}
\caption{\small The dipole and quadrupole polarizabilities.
 The numbers in brackets correspond 
to the order~$p^6$ LECs in Eq.~(\ref{eq:LECsp6bijnensprades}).}
\label{tab:polar1}
\vspace*{0.2cm}
\begin{tabular}{c||c|c}
\hline
    & to one loop & to two-loops \\ \hline
$(\alpha_1-\beta_1)_{\pi^+}  $ & $ 6.0  $ & $ 5.7 \,\, [5.5]    $
\\ 
$ (\alpha_1+\beta_1)_{\pi^+} $ & $ 0    $ & $ 0.16\,\, [0.16]   $
\\ \hline
$ (\alpha_2-\beta_2)_{\pi^+} $ & $ 11.9 $ & $ 16.2\,\, [21.6]   $ 
\\ 
$ (\alpha_2+\beta_2)_{\pi^+} $ & $ 0    $ & $-0.001\,\,[-0.001] $
\\ \hline
\end{tabular}
\end{center}
\end{table}

\subsection{Estimating the uncertainties}
\label{subsec:uncertainties}

The uncertainty in the prediction for the polarizability has two sources.
First, the low-energy constants are not known precisely. Second,
we are dealing here  with an expansion in powers of the momenta and of the
quark masses. We carried out this expansion up to and including
terms of order~$p^6$. Higher order terms will influence the result -
 by which amount?

We start the discussion by considering the  uncertainties in the LECs.
 We shall use the order $p^4$ LECs displayed 
in Eqs.~(\ref{eq:LECsp4}) and (\ref{eq:LECs65}).
The LECs at order~$p^6$ are not well known, see the discussion in 
Section~\ref{sec:effective}. In Table~\ref{tab:LECsp6}, 
we display the contributions from the LECs
at order~$p^6$ to the four polarizabilities, using the values
from Eq.~(\ref{eq:LECsp6}). The ones 
corresponding to Ref.~\cite{BijnensPrades}
 - displayed in Eq.~(\ref{eq:LECsp6bijnensprades}) -  
are given in square brackets. The only significant difference
 occurs in the value of the difference of the quadrupole 
polarizabilities $(\alpha_2-\beta_2)_{\pi^\pm}$.

\begin{table}[ht]
\begin{center}
\caption{\small The contribution of the LECs at order~$p^6$ to the
 polarizabilities, according to Eq.~(\ref{eq:LECsp6}). 
The numbers in brackets correspond to
Eq.~(\ref{eq:LECsp6bijnensprades}).}
\label{tab:LECsp6}
\vspace*{0.2cm}
\begin{tabular}{c||c|c|c||c}
\hline
    & $a_1^r$        &  $a_2^r$    &  $b^r$ & total
 \\ \hline
$(\alpha_1-\beta_1)_{\pi^\pm} $ & 
$ -0.14\,\,[-0.37] $ & $ 0 $ & $ 0.14\,\,[0.13] $  & $0\,\,[-0.24] $  \\
$ (\alpha_1+\beta_1)_{\pi^\pm} $ & 
$ 0 $    & $ 0 $ & $0.14\,\,[0.13] $ &$0.14\,\,[0.13] $ \\
$ (\alpha_2-\beta_2)_{\pi^\pm} $ & 
$ 0 $    & $0.72\,\,[6.09] $ &  $-0.83\,\,[-0.78] $ &$-0.10\,\,[5.31] $\\
$ (\alpha_2+\beta_2)_{\pi^\pm}$ & $0 $    & $0 $ & $0 $ &$0 $ \\
\hline
\end{tabular}
\end{center}
\end{table}
\vskip5mm

\begin{table}[ht]
\begin{center}
\caption{\small Resonance saturation induces a scale dependence
in the amplitudes. Displayed are the values of the polarizabilities
in case that saturation is assumed at $\mu=500$ MeV or at  $\mu=1$ GeV.}
\label{tab:scale}
\vspace*{0.2cm}
\begin{tabular}{c||c|c}
\hline
    & $\mu=500$ MeV        &  $\mu=1$ GeV \\
\hline
$ (\alpha_1-\beta_1)_{\pi^\pm} $ & $ 6.1 $ & $5.5  $   \\
$ (\alpha_1+\beta_1)_{\pi^\pm} $ & $ 0.20  $ & $0.13 $   \\
$ (\alpha_2-\beta_2)_{\pi^\pm} $ & $ 14.6  $ & $17.2  $   \\
$ (\alpha_2+\beta_2)_{\pi^\pm}$ & $-0.001 $ & $-0.001  $   \\
\hline
\end{tabular}
\end{center}
\end{table}

Resonance saturation generates a second inherent uncertainty -
 should one saturate at $\mu=500$ MeV or at $\mu=1$ GeV?
In Table~\ref{tab:scale} we display the polarizabilities
for saturation at these two scales, using Eq.~(\ref{eq:LECsp6}).
 It is seen that the induced change (which is independent of the values 
of the LECs at order~$p^6$) 
is  quite substantial for $(\alpha_2-\beta_2)_{\pi^\pm}$. This is
related to the fact that this quantity contains 
a rather substantial chiral logarithm, see below.

We now discuss the second source of uncertainties, the truncation of
the chiral expansion itself. It is clear that, to have an idea
of higher order terms, one needs at least the first two terms in the 
expansion. This makes it already clear that it is difficult
to make reliable predictions for the polarizabilities
connected with the helicity flip amplitude, from
which we have determined here the leading order contribution only.
So, let us concentrate first on the helicity non-flip case $H_{++}$.

The Born-term subtracted helicity amplitude
$H_{++}$ does not have branch points at the Compton threshold.
This is why it can be expanded there into an ordinary Taylor series e.g. 
in the variable $s$ and $\nu=(t-u)$.
One then expects that the amplitude 
 at the Compton threshold is less affected
by chiral logarithms than its counterparts at the threshold
for $\gamma\gamma\to\pi\pi$, where unitarity cusps do occur. This is 
illustrated in Fig.~\ref{fig:Hpp}, where we display 
the quantity $10^2 M_\pi^2H_{++}(s,t=u)$ as a function of $s$
at $t=u$. Above the threshold $s=4 M_\pi^2$, the modulus is shown. 
The solid (dashed) line is the expression to two loops
(to one loop). It is clearly seen that the corrections at  the Compton 
threshold are much smaller than the ones at the threshold
for $\gamma\gamma\to\pi\pi$. To identify the chiral logarithms, we note
that according to Eq.~(\ref{RGE}), the quantities ${\bar l}_i$ 
diverge in the chiral limit, ${\bar l}_i=-l + \Delta^r_i$, where $\Delta^r_i$
is quark mass independent. We decompose the ${\bar l}_i$ in the amplitude
$H_{++}$ in this manner and display the contributions from the chiral 
logarithms $l$ in Table~\ref{tab:chirallogH}. 
The numbers illustrate that, indeed, chiral logarithms
generate a smaller contribution at the Compton threshold than at threshold
for pion pair production.

\begin{figure*}[ht]
\begin{center} 
\epsfig{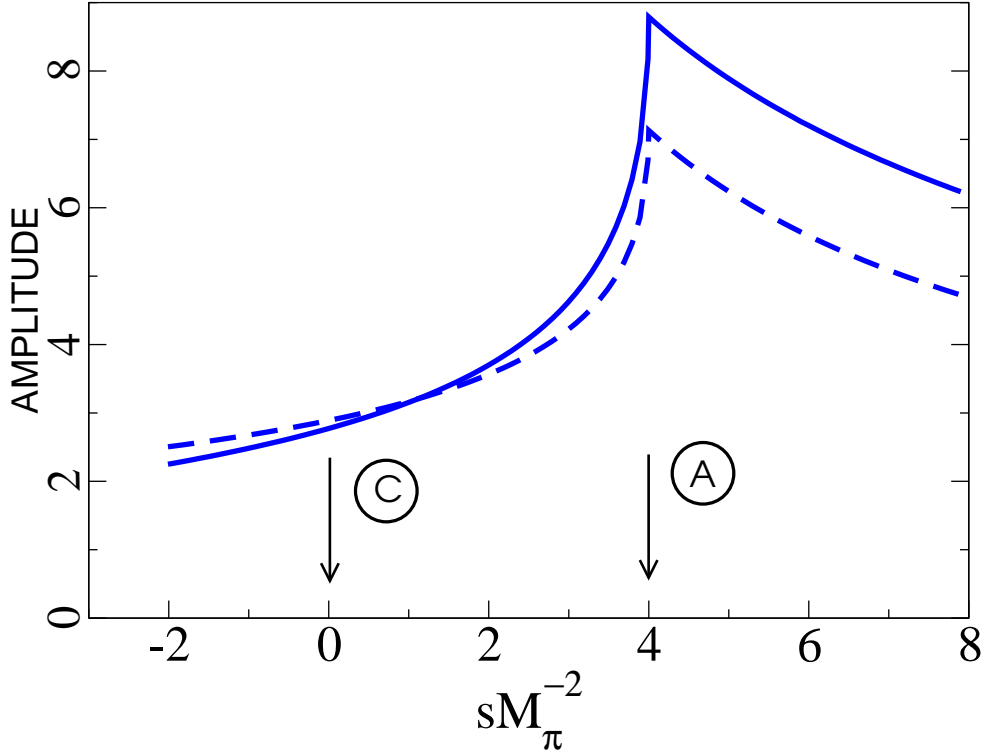} 
\caption{\small  The helicity non-flip amplitude $H_{++}$ in units
of $M_\pi^2$ as a function of $s$, at $t=u$, with Born term subtracted.
For $s\le 4 M_\pi^2$ the quantity shown is $10^2M_\pi^2 H_{++}$,
and for $s\ge 4 M_\pi^2$ we display $10^2M_\pi^2 |H_{++}|$.
The solid (dashed) line is the expression to two loops
(to one loop).
The Compton threshold in $\gamma\pi^\pm\to\gamma\pi^\pm$ and the
threshold in $\gamma\gamma\to\pi^+\pi^-$ are denoted by the encircled
letters $C$ and $A$, respectively. It is clearly
seen that two-loop corrections are suppressed at the Compton threshold.}
\label{fig:Hpp}
\end{center}
\end{figure*}
\begin{table}[t]
\protect
\begin{center}
\caption{\small 
The amplitude $10^2M_\pi^2H_{++}(s,t=u)$ at the Compton threshold $s=0$ 
and at $s=4M_\pi^2$. The contribution from
chiral logarithms, listed in the fourth  column, is
included in the two-loop result quoted in column three. The normalization is 
$N_H=10^2M_\pi^2$. The LECs at order~$p^6$ are the 
ones from Eq.~(\ref{eq:LECsp6}).}\label{tab:chirallogH}
\vspace{1em}
\begin{tabular}{c||c|c|c} \hline
$N_HH_{++}(s,t=u)$  & to 1 loop & to 2 loops &chiral logarithms  \\ 
\hline
$s=0$          & $ 2.89 $ & $ 2.77 $ & $- 0.35 $ \\ 
\hline
$s=4M_\pi^2$   & $ 7.13 $ & $ 8.80 $ & $ 1.28 $  \\ 
\hline
\end{tabular}
\end{center}
\end{table}
 As for the quadrupole polarizabilities $(\alpha_2-\beta_2)_{\pi^\pm}$,
also connected with the non-flip amplitude, 
it is seen from Table~\ref{tab:polar1} that there is a substantial
two-loop correction to the one-loop result.
 This can be again understood from the behavior of $H_{++}$.
As its value is considerably increased at the threshold 
$\gamma\gamma\to\pi\pi$, its slope at the Compton threshold
receives a substantial correction as well, in order to make up
that change. Indeed, the chiral expansion generates chiral logarithms
that are responsible for the major part of the increase.
If we decompose $(\alpha_2-\beta_2)_{\pi^\pm}$
in a manner analogous the amplitude $H_{++}$ above,
we find that chiral logarithms contribute
 with  $\simeq 4.5\times 10^{-4}\,{\rm fm}^5$
at two-loop order. These logarithms are, of course,
independent of the LECs at order~$p^6$.

Finally, we have checked whether  there are potentially large
contribution to $H_{++}$ at order~$p^8$. Using the same procedure as in
\cite{GIS}, we found that all contributions from resonance exchange 
with masses below 1 GeV have a negligible effect - we do not quote the 
corresponding numbers here.

We now turn to the helicity flip amplitude $H_{+-}$, which starts out 
at order~$p^6$: we have determined here only its leading order term 
in the chiral expansion. 
We checked whether there are potentially large
contribution to $H_{+-}$ at order~$p^8$, as is the case in
$\gamma\gamma\to \pi^0\pi^0$ \cite{GIS}. Whereas, there is substantial
contribution from  $\omega$-exchange in the neutral case, this resonance
does not contribute here, and the remaining contributions from 
$\rho$-exchange are very small, except for the contribution to 
the slope, parametrized by $(\alpha_2+\beta_2)_{\pi^\pm}$,
 which is affected by $-0.04 \times 10^{-4}$ fm$^5$.
On the other hand,  there are
also contributions from one-loop graphs at order~$p^8$, 
where each vertex is generated by an
anomalous contribution $\gamma\pi\to\pi\pi$ 
from the Wess-Zumino-Witten Lagrangian
at order~$p^4$. We see no reason why these should be small
compared to the leading term at order~$p^6$. We, therefore, do not consider
the chiral prediction for these quantities particularly reliable.   

\subsection{Values of the polarizabilities}
We have now all ingredients to provide a value for the 
 polarizabilities in the chiral expansion, and start the discussion with 
$(\alpha_1-\beta_1)_{\pi^\pm}$.
We add the uncertainties from the couplings at order~$p^4$, from $b^r$ and 
from the scale dependence introduced by the resonance scheme 
in quadrature and obtain
\be\label{eq:uncdipole}
\Delta=0.80\times 10^{-4}\,{\mbox{fm}}^3\,
\ee
for the so generated  uncertainty.
\begin{figure}[!hbp]
\begin{center}
\epsfig{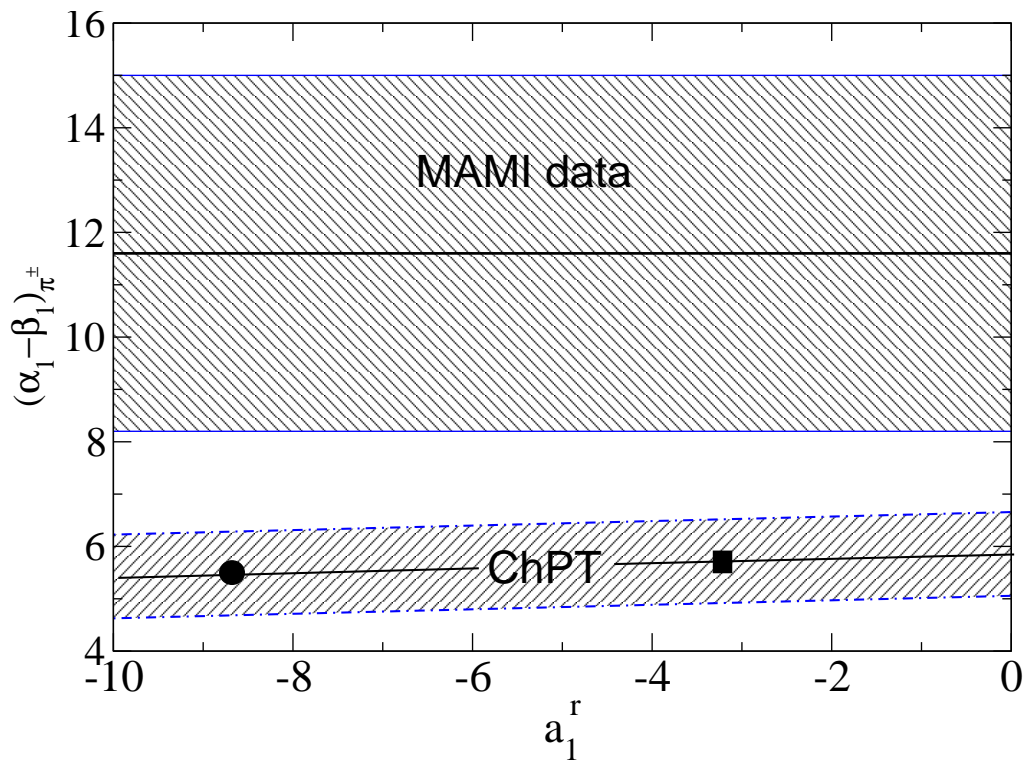}
\end{center}
\caption{\small The polarizabilities $(\alpha_1-\beta_1)_{\pi^\pm}$ 
at two-loop order, as a function of $a_1^r$. 
 The filled in square [filled in circle]  denotes the value obtained by using
 for $a_1^r$ the value from Eq.~(\ref{eq:LECsp6})
 [(\ref{eq:LECsp6bijnensprades})].
 The width of the ChPT band is calculated in the manner described 
in the text. ``MAMI data'' refers to 
the recent MAMI experiment  \cite{MAMI}. 
We have added the errors quoted there in quadrature.}
\label{fig:amb1}
\end{figure}
We  display in 
Fig.~\ref{fig:amb1} the result as a
function of the rather inaccurately known coupling $a_1^r$.
We indicate with a filled square [filled circle] the 
two-loop result, evaluated with the LECs given in Eq.(\ref{eq:LECsp6}) 
[\ref{eq:LECsp6bijnensprades}]. The width of the band is twice the
uncertainty displayed in Eq.~(\ref{eq:uncdipole}). 
Let us note that the two-loop prediction differs only slightly from 
the one-loop calculation, see Table~\ref{tab:polar1}. 
This again shows that the value for the dipole polarizability is
rather reliable - there is no sign of any large, uncontrolled  correction 
to the two-loop result. We use the maximum deviation $1.0$ from
the central value $5.7$ as the final
theoretical uncertainty for the dipole polarizability, 
and obtain
\bea\label{eq:polarfinal}
(\alpha_1-\beta_1)_{\pi^\pm}= (5.7\pm 1.0)\times 10^{-4}\,{\rm fm}^3\,.
\eea
In the same figure, we also
show the result of the recent measurement at MAMI \cite{MAMI} of 
this quantity. It is seen that the chiral expansion is in conflict with
 that measurement, independent of any reasonable value chosen for $a_1^r$.
 The discrepancy with the recent 
dispersive analysis by Fil'kov and Kashevarov 
\cite{FK05}, displayed in  Table~\ref{tab:polar3}, is even more 
pronounced.

\begin{table}[ht]       
\centering
\caption{\small 
Experimental information on $(\alpha_1-\beta_1)_{\pi^\pm}$,
 in units of $10^{-4}\,{\rm fm}^3$.
 We indicate the reaction and the data used. 
In \cite{DH93} and \cite{Babusci}, $\alpha_1$ was determined, using as a 
constraint $\alpha_1=-\beta_1$. To obtain $(\alpha_1-\beta_1)_{\pi^\pm}$, we
multiplied the results by a factor of 2.}
\label{tab:polar3}
\vspace*{0.2cm}
\def\arraystretch{1.2}
\begin{tabular}{ll|c} 
\hline
\multicolumn{2}{l|}{Experiments}    & $(\alpha_1-\beta_1)_{\pi^\pm}$  \\ 
\hline
$\gamma p\to \gamma \pi^+ n$ & Mainz (2005) \cite{MAMI} & 
$11.6\pm 1.5_{\,\rm stat}\pm 3.0_{\,\rm syst} \pm 0.5_{\,\rm mod}$ \\
\hline
\multicolumn{2}{l|}{L. Fil'kov, V. Kashevarov (2005) \,\cite{FK05}}   & 
 $13.0^{+2.6}_{-1.9}$  \\ 
 $\gamma\gamma\to\pi^+\pi^-$ & MARK II \,\cite{MARK-II}\,, & \\ 
\multicolumn{2}{l|}{ TPC/2$\gamma$ \cite{TPC}\,,\,   
 CELLO \,\cite{CELLO}\,, } & \\
\multicolumn{2}{c|}{VENUS\, \cite{VENUS}\,,   
 ALEPH \,\cite{ALEPH}\,,\, BELLE \cite{BELLE} }& \\
\hline
\multicolumn{2}{l|}{ A. Kaloshin, V. Serebryakov (1994)\, \cite{KS94}} & 
$5.25\pm 0.95$\\ 
$\gamma\gamma\to\pi^+\pi^-$ & MARK II \,\cite{MARK-II}  & \\
&Crystal Ball Coll. \cite{crystalball}&\\
\hline
\multicolumn{2}{l|}{ J.F. Donoghue, B. Holstein (1993)\, \cite{DH93}  }  
& $5.4$ \\
$\gamma\gamma\to\pi^+\pi^-$ & MARK II\,\cite{MARK-II}  &       \\
\hline
\multicolumn{2}{l|}{D. Babusci {\em et al.} (1992)\, \cite{Babusci} } & \\
 $\gamma\gamma\to\pi^+\pi^-$ &   PLUTO \,\cite{PLUTO}  & 
 $38.2 \pm 9.6\pm 11.4$     \\
                             &    DM 1 \,\cite{DM1}  & $34.4\pm 9.2$  \\
                             &    DM 2 \,\cite{DM2} & $52.6\pm 14.8$ \\
                             & MARK II \,\cite{MARK-II} & $4.4\pm 3.2$\\
\hline
$\gamma p\to \gamma \pi^+ n$ &
Lebedev Inst. (1984) \cite{Lebedev} & $ 40\pm 24 $ \\
\hline
 $\pi^-Z\to\gamma\pi^-Z$ & Serpukhov (1983)\,\cite{Serpukhov} & 
$15.6\pm 6.4_{\,\rm stat}\pm 4.4_{\,\rm syst}$ \\
\hline
\end{tabular}
\end{table}
Next, we discuss the quadrupole case $(\alpha_2-\beta_2)_{\pi^\pm}$. 
Here, the situation differs. First, as we discussed above,
chiral logarithms at order~$p^6$ 
generate now a rather large contribution. As a result of this, the scale
dependence of the saturation scheme is  pronounced as well, see 
Table~\ref{tab:scale}. Second, the LEC $a_2^r$ enters with weight 12.
 As a result of this, 
it now matters which value of $a_2^r$ is used,
even though its contribution is suppressed by the 
factor ${{M_\pi^2}(16\pi^2  F_\pi^2)^{-1}\simeq 0.014}$.
Although the
effect of using the \cite{BijnensPrades} value for $a_2^r$ goes 
in the right direction and brings the result in closer agreement
with the analysis of Fil'kov and Kashevarov \cite{FK05},
\bea\label{eq:filkov}
(\alpha_2-\beta_2)_{\pi^\pm}=(25.0^{+0.8}_{-0.3})\times 10^{-4}\,{\rm fm}^5\,\,,
\eea
we cannot take the outcome as a reliable two-loop prediction of ChPT:
 compared to the one-loop result, the two-loop contribution 
 generates nearly a 100 percent contribution in this case, see 
Table \ref{tab:polar1}.

As we discussed in the previous subsection, the situation is no better for
the polarizabilities $(\alpha_{1,2}+\beta_{1,2})_{\pi^\pm}$ associated with
the helicity flip amplitude: considerably more work is required to put
the chiral prediction on a firm basis. Orders of magnitudes
can be read off from Table~\ref{tab:polar1}.

In Table~\ref{tab:polar3}, we collect 
 experimental information 
on the dipole polarizabilities of the charged pions. In view of the large
scatter of the results, which are partly inconsistent among each other, 
it is interesting to note that the COMPASS
collaboration at CERN is presently remeasuring pion and kaon 
polarizabilities with high statistics, using the
Primakoff effect \cite{COMPASS}. It is also worth mentioning 
that a reanalysis of the MAMI data \cite{MAMI}, using a chiral invariant
framework, is underway \cite{scherer}.

\setcounter{equation}{0}
\section{Summary and outlook}

\begin{enumerate}
\item[1.]
We have recalculated the two-loop expression for the
amplitude $\gamma\gamma\rightarrow\pi^+\pi^-$ in the framework of chiral
perturbation theory. We have made use of
 the   techniques developed in Ref.~\cite{GIS,GS}, and of the
 effective Lagrangian ${\mathcal L}_6$ constructed in \cite{BCE1,BCE2}.
\item[2.]
The two Lorentz invariant amplitudes $A$ and $B$  are
 presented as a sum over multiple Feynman
parameter integrals, whose numerical evaluation poses no difficulty.
This is in contrast to Ref.~\cite{Burgi}, where part of the
amplitudes, denoted by $\Delta_{A,B}$,  could be published in 
numerical form only. Further, the method has allowed us to evaluate
the dipole and quadrupole polarizabilities in closed form.
As far as we are aware, the quadrupole polarizabilities for charged
pions have never been calculated in \chpt before.
\item[3.]
Our result agrees with the earlier calculation \cite{Burgi} 
up to the remainder $\Delta_{A,B}$. The induced changes in  
the cross section and in the dipole polarizabilities are far below 
the uncertainties generated by the (not precisely known) 
values of the low-energy constants. For the cross section below 500 MeV,
the change is less than 1 percent. For the polarizabilities, the change
is given in columns two and three of Table~\ref{tab:comparisonpol}.
\item[4.]
We have investigated the uncertainties in the polarizabilities
due to higher order corrections, and due to  the uncertainties
in the LECs at order~$p^6$. According to our analysis,
the two-loop result Eq.~(\ref{eq:polarfinal}) for the dipole polarizability 
 $(\alpha_1-\beta_1)_{\pi^\pm}$
 is  particularly reliable. It is in conflict with the 
recent experimental result obtained at MAMI \cite{MAMI}, 
or with the dispersive analysis performed in \cite{FK05}.
\end{enumerate}

\ack
This work was completed while M.A.I. visited the University of Bern. 
It is a pleasure to thank Lev Fil'kov for useful  correspondence.
This work was  supported  by the Swiss
National Science Foundation, by RTN, BBW-Contract No. 01.0357,
and EC-Contract  HPRN--CT2002--00311 (EURIDICE).
M.A.I. also appreciates the partial support by
the Russian Fund of Basic Research under Grant No. 04-02-17370.  
Also, partial support by the Academy of Finland, grant 54038,
is acknowledged.

\appendix

\setcounter{equation}{0}
\section{Comparison with the previous calculation}
\label{app:comparison}
In this appendix, we compare the amplitudes $A,B$ and the dipole
polarizabilities  
 with the previous calculation
 presented in  Ref.~\cite{Burgi}. In that reference, the 
amplitudes were evaluated  with a
different technique. Furthermore,  the Lagrangian ${\mathcal L}_6$
was not available in those days, and an important 
ingredient to check the final result was, therefore,  missing. 
We can make the following observations.
\begin{enumerate}
\item
The amplitudes $A$ and $B$ consist of a part with explicit 
analytic expressions, and additional terms $\Delta_{A,B}$, 
that are displayed in Appendices~\ref{app:deltaab} and 
\ref{app:polynomials} of the present work in the 
form of integrals over Feynman parameters.
$\Delta_{A,B}$ were given only in numerical form in \cite{Burgi}.
\item
We find that our explicit analytic expressions agree with the ones in 
\cite{Burgi}.
To compare $\Delta_{A,B}$, we  made two checks. First, we 
 evaluated the cross section  for the reaction 
 $\gamma\gamma\rightarrow\pi^+\pi^-$ below a centre-of-mass
energy of 500 MeV, using the same values for the LECs 
as in \cite{Burgi},
\bea\label{eq:LECsburgi}
(\bar l_1,\bar l_2,\bar l_3, \bar l_4,\bar l_\Delta)&=&
(-1.7,6.1,2.9,4.3,2.7)\,,\nnnl
(a_1^r,a_2^r,b^r)&=&
(-3.3,0.75,0.45)\,.
\eea
Our result agrees within a fraction of a percent with the two-loop 
result displayed in Fig.~9 of Ref.~\cite{Burgi}.
\item
Second, we  evaluated the dipole polarizabilities 
by using the values of the LECs in Eq.~(\ref{eq:LECsburgi}).
The result is displayed in the third column  of  
Table~\ref{tab:comparisonpol}. In the second column, 
we display the values obtained in \cite{Burgi}. It is seen that 
the two results agree very well. The small difference 
is entirely due to the 
slightly different values of the quantities $\Delta_{A,B}$.
\end{enumerate}
\begin{table}[!ht]
\begin{center}
\caption{\small Comparison of the polarizabilities with the previous
  calculation by Burgi \cite{Burgi}. The second column displays Burgi's
  result, the third one our evaluation, using Eq.~(\ref{eq:LECsburgi}).
The fourth column displays the polarizabilities
evaluated with the LECs used in the present work.}
\label{tab:comparisonpol}
\vspace*{0.2cm}
\begin{tabular}{c||c|c|c}
\hline
    & Burgi \cite{Burgi}& Present work&Present work\\
&&LECs from Eq.~(\ref{eq:LECsburgi})&LECs from 
Eqs.~(\ref{eq:LECsp4}) - (\ref{eq:LECsp6})\\ \hline
$(\alpha_1-\beta_1)_{\pi^+}$ &$ 4.42  $ & $ 4.39$&$5.72$\\ 
$ (\alpha_1+\beta_1)_{\pi^+} $ & $0.31 $ &$0.28$  & $ 0.16$\\ \hline
\end{tabular}
\end{center}
\end{table}

The following comments concerning the central values of
 the polarizabilities are in order. We display in the last column in Table
 \ref{tab:comparisonpol} the values of polarizabilities 
obtained by using the LECs from Eqs.~(\ref{eq:LECsp4}) - (\ref{eq:LECsp6}). 
It is seen that the combination
$(\alpha_1-\beta_1)_{\pi^\pm}$ differs considerably from the 
one obtained with Eq.~(\ref{eq:LECsburgi}). To identify the source of this
difference, we first note that the polarizabilities depend 
linearly on the LECs, except the term $\bar l_\Delta \bar l_4$ in $d_{1-}$,
as can be seen from the Eqs.~(\ref{eq:alphabeta}-\ref{polchi4}).
We then expand $(\alpha_1-\beta_1)_{\pi^\pm}$ in the LECs around 
their central values, 
drop  quadratic terms  and obtain at
 $\mu=M_\rho$ the decomposition 
\bea\label{eq:comparisonLECsp4}
(\alpha_1-\beta_1)_{\pi^\pm}&=&
5.719\nnnl
&+&0.215\cdot(\bar l_1+0.4)-0.177\cdot(\bar l_2-4.3)
  -0.014\cdot(\bar l_3-2.9)\nnnl
&+&0.172\cdot(\bar l_4-4.4)+2.44\cdot(\bar l_\Delta-3)+0.04\cdot(a_1^r+8b^r)\, .
\eea
Inserting in this expansion e.g. the values 
in Eq.~(\ref{eq:LECsburgi})
 generates the result in column 3 of Table \ref{tab:comparisonpol}.
It turns out that  all changes in the order~$p^4$ LECs 
 have conspired to modify Burgi's result for $(\alpha_1-\beta_1)_{\pi^\pm}$ 
towards positive values. 
The contribution from the LECs at order $p^6$ is negligible.

In summary, we conclude that our calculation  nicely 
confirms Burgi's result \cite{Burgi}, provided that the same set of LECs is
used.

\section{One-loop functions}
\label{app:loopfunc}

In order to simplify the expressions, we set the pion mass equal 
to one in the following Appendices,
\bea
M_\pi=1\, .
\eea

1. The loop-integral $\bar{G}(s)$ is
\be
         \bar{G}(s) = - \frac{1}{16 \pi^2} \left \{ 1 + \frac{2}{s}
         \int^{1}_{0} \frac{dx}{x} \ln (1-s\,x(1-x)) \right \} \; .
         \ee
$\bar{G}$  is analytic in the complex $s$ - plane, cut along the positive real
axis  for Re~$ s \geq 4$. At small $s$,
\be
\bar{G}(s)= \frac{1}{16\pi^2}\sum_{n=1}^\infty s^n
\frac{(n!)^2}
{(n+1)(2n+1)!}\;.
\ee
         The absorptive part is
         \bea
         \mbox{Im}\, \bar{G} (s) &=& \frac{1}{8\,s \,\pi }
         \ln\left\{
\frac{1+\sigma}{1-\sigma}\right\} \,, \hspace{1cm}  s>4 \,, \nnnl
        \sigma &=& \sqrt{1-4/s}\, ,
         \eea
and
        \be
          -  16 \pi^2 \bar{G}(s)  =  \left\{
         \begin{array}{lll} 1 & + \frac{1}{s} \left(\ln
         \frac{1-\sigma}{1+\sigma} + i \pi \right)^2 ; &  \qquad 4  \leq s \,,
         \\
         1 & - \frac{4}{s} \mbox{arctg}^2 (\frac{s}{4-s})^\frac{1}{2}\, ;&\qquad
 0   \leq s \leq 4\,,
         \\
 1 & + \frac{1}{s} \ln^2 \frac{\sigma - 1}{\sigma+1}\,; & \qquad s \leq 0\,.
         \end{array} \right.
         \ee

         In the text we also need
         \bea
         \stackrel{=}{G} (s) &=& \bar{G} (s) - s\,\bar{G}'(0) \,.
         \eea
2. The loop-integral $\bar{J}(s)$ is
\be
\bar{J}(s) = -\frac{1}{16\pi^2}\int_0^1 dx \ln(1-s\,x(1-x)) \; .
\ee
$\bar{J}$ is analytic in the complex $s$ - plane, cut along the positive real
axis  for Re~$ s \geq 4$. At small $s$,
\be
\bar{J}(s)= \frac{1}{16\pi^2}\sum_{n=1}^\infty  s^n
\frac{(n!)^2}
{n(2n+1)!}\;.
\ee
The absorptive part is
\be
\mbox{Im} \,\bar{J}(s) = \frac{\sigma}{16\pi}\,,\hspace{1cm} s>4 \; .
\ee
Explicitly,
\bea
          16 \pi^2 \bar{J}(s)  =  \left\{
         \begin{array}{lll} &
          \sigma \left(\ln \frac{1-\sigma}{1+\sigma}+i \pi \right) +2\,; &
 \qquad 4  \leq s \,,
         \\
          &2-2(\frac{4-s}{s})^\frac{1}{2} \mbox{arctg}
(\frac{s}{4-s})^\frac{1}{2}\,;
 &\qquad
0          \leq s \leq 4\,,
         \\
          & \sigma \ln \frac{\sigma - 1}{\sigma+1}+2 \,;& \qquad s  \leq
0\,.          \end{array} \right.
\eea
In the text we also need
         \bea
         \stackrel{=}{J} (s) &=& \bar{J} (s) - s\,\bar{J}'(0) \,.
         \eea
3. The loop-function $\bar{H}$ is defined in terms of $\bar{G}$ and
$\bar{J}$, \be
\bar{H}(s)=(s-10)\bar{J}(s) +6\, \bar{G}(s) \,.
\ee

\section{The quantities $\Delta_A$ and $\Delta_B$}
\label{app:deltaab}

Here we display the expressions for the quantities $\Delta_{A(B)}$.

\begin{eqnarray}
&&
\hspace*{-1cm}
\Delta_{A} (s,t,u) = 
\frac{1}{(4\pi F_\pi)^4 }
\left\{ \left( \frac{35947}{8100}+\frac{\pi^2}{3} \right)
       -\frac{1130291}{6350400} s -S(t)-S(u)\right\}
\nnnl
&&
\hspace*{1cm}
+ \frac{1}{(4 \pi F_\pi)^4} \frac{1}{288}  
\left\{   F_A^{\rm acn}(  s,  t,  u) 
       + F_A^{\rm ver}(  s)
       + F_A^{\rm box}(  s,  t,  u)\right\},
\label{delA}\\
\nnnl
&&
\hspace*{-1cm}
\Delta_{B} (s,t,u) = 
\frac{1}{(4 \pi F_\pi)^4 }
\left\{ \frac{492197}{1411200}
-\left( \frac{81101}{70560}-\frac{\pi^2}{6}
       +\frac{1}{2}S(t) +\frac{1}{2}S(u) \right)\frac{1}{s}
\right\}
\nnnl
&&
\hspace*{1cm}
+ \frac{1}{(4 \pi F_\pi)^4} \frac{1}{576} 
     \left\{F_B^{\rm acn}(  s,  t,  u) 
         + F_B^{\rm ver}(  s)
        + F_B^{\rm box}(  s,  t,  u)\right\} . 
\label{delB}
\end{eqnarray}
Here, the function $S(t)$ is defined by

\begin{eqnarray}
\label{sunset}
S(t) & = & \frac{h_F(t)}{6 (t-1)^2}\, ,
\\
&&\nnnl
h_F(t) & = &-  \int\limits_4^{\infty} d\sigma \beta(\sigma) \int_0^1 dx
\left\{ 6 t +(1-12 x+18 x^2)  t^2 \right\} 
\nnnl
&& 
\times \left\{
\ln \frac{z_2(x,t)}{z_2(x,1)}+\frac{(t-1) x  (1-x)}{z_2(x,1)}
\right\} \, , \nn \\[0.2cm] 
\beta(\sigma) &=& \sqrt{1-4/\sigma}\, ,\nn \\[0.1cm] 
z_2(x,t) & = & x^2+\sigma (1-x) -(t-1) x (1-x)\,.
\nn
\end{eqnarray}
The loop function $h_F(t)$ stems from the sunset diagram.
The functions $F_I$ are stemming from
the acnode, vertex and box diagrams and are defined by
 
\begin{eqnarray}
F_{I}^{\rm acn} &=& \int\limits_4^\infty\! d\sigma \beta\!
                    \int\limits_0^1\! d^3x
\left\{
 \frac{P_{I;\, {\rm acn}(1)}} {y\cdot z_{{\rm acn}(1)}}
+\sum\limits_{i=2}^6 \frac{P_{I;\, {\rm acn}(i)}} {z^2_{{\rm acn}(i)}}
\right\} \,,
\nnnl
&&\nnnl
F_{I}^{\rm ver} &=& 
\int\limits_4^\infty\! \frac{d\sigma \beta}{\sigma}\!
\int\limits_0^1\! d^2x
\cdot\frac{P_{I; {\rm ver}}}{z_{\rm ver}}\,  ,
\nnnl
&&\nnnl
F_{I}^{\rm box} &=& 
\!\!
\int\limits_4^\infty\! \frac{d\sigma \beta}{\sigma}
\int\limits_0^1\! d^3x
\sum\limits_{n=1}^2
\left\{
 P_{I;  \rm{box_+}}^{(n)} D^{(n)}_{\rm box_+} 
+P_{I;  \rm{box_-}}^{(n)} D^{(n)}_{\rm box_-}
\right\}\,, \quad   I=A,B  ,
\eea
and
\bea
D^{(n)}_{\rm box_\pm} &=&
\frac{1}{z^n_{\rm box; t}} \pm \frac{1}{z^n_{\rm box; u}} \, .
\nn
\end{eqnarray}
Here $P_{I}$  are polynomials in 
 $s, \nu=t-u$ and in $x_i$. Their explicit expressions are given in 
Appendix~\ref{app:polynomials}. The arguments of the functions
are defined by
\bea
z_{{\rm acn}(i)} & = & y -a_i x_3 (1-x_3)\,, \qquad (i=1,\ldots, 6)\,,
\qquad y=x_3^2+\sigma (1-x_3)\,,
\nnnl
a_1 &=& a_2 = x_1 x_2 s + x_1 (t-1) +x_2 (u-1)\,,
\nnnl
a_3 &=& t-1\,, \quad a_4 = x_1 (t-1)\,, 
\nnnl
a_5 &=& u-1\,, \quad a_6 = x_2 (u-1)\,,
\nnnl
z_{\rm ver} &=& \sigma (1-x_3) +x_3^2  y_2\,,
\hspace{1cm} y_2 = 1-s x_2 (1-x_2)\,,
\nnnl
z_{\rm box; t} &=& B_{ \rm t}-A_{ \rm t} x_1 \,,
\nnnl
A_{ \rm t} &=& x_2 x_3 \Big[s (1-x_2) x_3 + (1-t) (1-x_3)\Big]\,,
\nnnl
B_{ \rm t}&=&A_{ \rm t} + z_{\rm ver}\, ,\nnnl
z_{\rm box; u}&=&z_{\rm box; t}|_{t\rightarrow u}\, .
\eea
The acnode integrals are easy to evaluate numerically
in the physical region for the reaction $\gamma\gamma\to\pi\pi$, because
 branch points occur at $t= 4,u=4$ only.
On the other hand, the vertex and box integrals contain branch 
 points at $s=4$.
 In order to evaluate these integrals at $s\ge 4$, we invoke dispersion
 relations in the manner described in \cite{GS}.

\section{The polynomials $P_A$ and $P_B$}
\label{app:polynomials}
Here, we display the polynomials $P_{A(B)}$ that occur in the expressions
$\Delta_{A(B)}$ in Appendix~\ref{app:deltaab}. We use the abbreviations
\bea
x_{+} &=& x_1+x_2-2  x_1  x_2 \,,\qquad  x_{-} = x_1-x_2\,  ,\nnnl
x_{123}&=&(1+x_3-2 x_2  x_3)(1-x_3+2 x_1 x_2 x_3)\,  .
\eea

\subsection{The polynomials $P_A$}

\begin{eqnarray*}
P_{{\rm A;\,  acn}(1)} &=& \;\;\; 144 x_3 (1-x_3)(s x_{+} -\nu  x_{-}), 
\\
&&\\
P_{{\rm A;  acn}(2)} &=&\;\;\;
12 s^2 (1-x_3)^2 \Big\{ x_+ (5 x_+ + 6 x_-^2-1) -5 x_+^2 x_3
-6 x_-^2 x_3^2
\\
&&
 + \Big[x_-^2 (9-6 x_+)-x_+ (2-7 x_+)\Big] x_3^3
\\
&&
+ \Big[6 x_-^2 (x_+ - 1)+(3-7 x_+) x_+\Big] x_3^4\Big\}
\\
&&
+ 12 s \nu (1-x_3)^2 x_- \Big\{
1-6 x_-^2-14 x_+ +10 x_+ x_3
\\
&&
+ 6 (1+2 x_-^2-x_+) x_3^2 - 2 (6 x_-^2+5 x_+ - 1) x_3^3
\\
&&
+ (6 x_-^2 + 8 x_+ -3 ) x_3^4 \Big\}
\\
&&
- 12 \nu^2 (1-x_3)^2 \Big\{6 x_+ (1-2 x_+) x_3^2
\\
&&
- x_-^2 \Big[ 9-5 x_3 - 12 x_+ x_3^2-(6-18 x_+) x_3^3
 + (5 - 12 x_+) x_3^4\Big]\Big\}
\\
&&
+ 24 \nu (1-x_3) x_- \Big\{
1+2 x_+ - (25+2 x_+) x_3 + 3 (7-8 x_+) x_3^2
\\
&&
+ (100 x_+ - 41) x_3^3 +(51-118 x_+) x_3^4 +(42 x_+ - 19) x_3^5\Big\}
\\
&&
- 12 s (1-x_3) \Big\{
2 x_+ (1+2 x_+) +  2 (2-12 x_-^2-17 x_+-2 x_+^2) x_3
\\
&&
+2 (13 - 6 x_-^2-11 x_+) x_3^2
- 2 (41-18 x_-^2 - 89 x_+ + 8 x_+^2) x_3^3
\\
&&
+ (82 - 36 x_-^2-194 x_+ + 28 x_+^2) x_3^4
\\
&&
-2 (15 - 6 x_-^2 - 35 x_+ + 6 x_+^2) x_3^5 \Big\}
\\
&&
+ 48 x_3 (1-x_3) \Big\{-2-4 x_+ +(35-50 x_+) x_3
\\
&&
+71 (2 x_+ - 1) x_3^2 + (61 -130  x_+) x_3^3
+ 2 (21 x_+ - 10) x_3^4 \Big\}\,,
\\
&&\\
P_{{\rm A;\,  acn}(3)} &=& \;\;\;
 3 (s-\nu)^2 (1 - x_3)^5 (1 + x_3)
\\
&&
+ 6 (s-\nu) (1-x_3)^3 (3+5 x_3+2 x_3^2-2 x_3^3 )
\\
&&
+ 12 x_3 (1-x_3) (2-x_3) (3 + 3 x_3 + x_3^2- x_3^3)\,,
\\
&&\\
P_{{\rm A;\,  acn}(4)} &=&\;\;\;
 24 (s-\nu)^2 x_1^2 (1-2 x_1)(1-x_3)^4 (2+x_3)
\\
&&
+ 48 (s-\nu) x_1 (1-2 x_1) (1-x_3)^2 (2+4 x_3-3 x_3^2)
\\
&&
+ 96 (1-2 x_1) x_3 (1-x_3) (2-x_3) (2+2 x_3-x_3^2)\,,
\\
&&\\
P_{{\rm A;\,  acn}(5)} &=& \;\;\;P_{{\rm A;\,  acn}(3)}|_{\rm t\leftrightarrow u}\,,
\\
&&\\
P_{{\rm A;\,  acn}(6)}&=&\;\;\;
P_{{\rm A;\,  acn}(4)}|_{\rm t\leftrightarrow u,\,x_1\leftrightarrow x_2}\,,
\\
&&\\
&&\\
P_{\rm A;\,  ver} &=&
 - 8 s^2 (1-2 x_2) x_2^2 x_3^4 \Big\{
-54+72 (3+4 x_2) x_3
\\
&&
+ (330 -1134 x_2-811 x_2^2) x_3^2
- 108 (5 - 15 x_2^2-4 x_2^3) x_3^3
\\
&&
- 45 x_2 (18-75 x_2+32 (3-x_2) x_2^2) x_3^4\Big\}
\\
&&
+ 32 s (1-2 x_2) x_2^2 x_3^4 \Big\{
33 +8 (27-7 x_2) x_3 
\\
&&
- 3 (222-30 x_2 +55 x_2^2) x_3^2 - 540 (1-3 x_2+x_2^2) x_3^3
\\
&&
+1215 (1-x_2)^2 x_3^4\Big\}\,,
\\
&&\\
&&\\
P_{\rm A;\,  box; +}^{(1)} &=&\;\;\;
4 s^2 x_2^2 x_3^3  
\Big\{
6 x_1 \Big[9 - 2 (23 + 8  x_2)  x_3 -  (67 - 405 x_2 + 31  x_2^2)  x_3^2
\\
&& 
+  (70 +  39 x_2 - 808 x_2^2 + 20 x_2^3)  x_3^3 
\\
&&
+  9  (6 - 61  x_2 + 53 x_2^2 + 60  x_2^3 )  x_3^4 
- 81  x_2  (3 - 10  x_2 + 8  x_2^2 )  x_3^5\Big]
\\
&& 
+  3  x_1^2  x_2  x_3  \Big[-92 - 171 x_3 + 592  x_2 x_3 
\\
&&
+ (231 + 4  (194 - 339  x_2)  x_2)  x_3^2 
\\
&&
+ 9  (1 - 2  x_2)  (71 - 22  x_2 - 60  x_2^2)  x_3^3 
\\
&&
+ 27  (1 - 2  x_2)^2  (11 - 16 x_2)  x_3^4 \Big] 
-  2  x_1^3  x_2^2  x_3^2  \Big[245 + (1 - 2  x_2)  x_3  
\\
&&
\times( 470 
+ 27  (1 - 2  x_2)  x_3  (15 + 8  (1 - 2  x_2)  x_3))\Big] 
\\
&&
+ 6  x_3  (-19 +24  x_3+ 35  x_3^2 - 36 x_3^3
\\
&&
+ x_2^2  x_3  (-25 + 40  x_3 + 27  (5 - 6  x_3)  x_3^2) 
\\
&&
+ x_2  (10 + 44 x_3  -199 x_3^2   + 9  x_3^3  (8 + 9  x_3) ) )
\Big\}
\\
&&
+ 12 \nu^2  x_2^3  x_3^4  
\Big\{ -20 - 16  x_1  (2 + 7  x_1)  
\\
&&
 +  \Big[48+50  x_2 + 310 x_1 + 101  x_1^2
 +  30 x_1  (3 + 15 x_1 -5 x_1^2) x_2 \Big]  x_3 
\\
&&
+ \Big[-78 + 2  x_1 + 411  x_1^2 
- 20  (9 + 63 x_1 + 57 x_1^2 + 13  x_1^3)   x_2 
\\
&&
+  120  x_1  (1 + x_1 + 5  x_1^2)  x_2^2\Big]    x_3^2 
-  3  \Big[24 + 3 x_1  (94 + 73  x_1) 
\\
&&
-  2  (69 + 351 x_1 + 381 x_1^2 + 19 x_1^3 )  x_2 
\\
&&
+  4  x_1  (24 +117 x_1 + 34  x_1^2 )  x_2^2\Big]  x_3^3 
\\
&&
+  9  (1 - 2  x_2)  \Big[18 +  54 x_1 - 36 x_1 x_2 
+  x_1^2  (33 -  2  (9 + 8  x_1)  x_2)\Big]   x_3^4 \Big\}
\\
&&
-48 s  x_2  x_3^2  
\Big\{
-6 + \Big[18 + 20  x_2 - 31  x_1  x_2\Big] x_3 
\\
&&
-  \Big[29 +  49 x_2  + 4 x_2^2 
   +  4  x_1 x_2  (9 - (22 -8  x_1)  x_2)\Big]  x_3^2 
\\
&&
+  \Big[-6 + 202 x_2  - 54  x_2^2  +164 x_1 x_2 
\\
&&
+  114 x_1 x_2^2 - 4 x_1  x_2^3 
- x_1^2 x_2^2 (149 -58  x_2) \Big] x_3^3 
\\
&&
+  \Big[65   -333 x_2 - 40  x_2^2 + 189 x_1 x_2
  -4 x_1  (137 + 75  x_2) x_2^2 
\\
&&
+  5  x_1^2 x_2^2  (23 + 12  x_2  (5 +  x_2))\Big]  x_3^4 
\\
&&
-    9  \Big[4-46 x_2^2 - x_1^2 x_2^2 (1-2 x_2) (41+8 x_2)
\\
&&
+  x_1 x_2 (25+18 x_2-76 x_2^2)\Big]  x_3^5 
\\
&&
+ 81  x_2 (1 - 2  x_2)
 \Big[2 -3 x_1 +x_1  (2 + x_1)  x_2  
 - 2  x_1^2  x_2^2\Big]  x_3^6\Big\}
\\
&&
+ 96  x_2  x_3^2  \Big\{
-21 + \Big[36 + 65  (1 - x_1)  x_2\Big]  x_3 
\\
&&
+  \Big[23 - 8  (23 - x_1)  x_2 + 176  x_1  x_2^2\Big]  x_3^2
\\
&&
 -  2 \Big[6 - (83 + 37  x_1)  x_2 +  120  x_1  x_2^2\Big]  x_3^3 
\\
&&
-  \Big[29 +   10  (17 - 7  x_1)  x_2 - 100  x_1  x_2^2\Big]  x_3^4 
\\
&&
-  9  \Big[8 - (31 + x_1)  x_2 + 32  x_1  x_2^2\Big]  x_3^5 
+  81  (1 -\! 2 \! x_2)  (1 - 2  x_1  x_2)  x_3^6\Big\}\,,
\\
&&\\
&&\\
P_{\rm A;\,  box; -}^{(1)} &=&\;\;\;
-8 s \nu   x_2^2  x_3^3  
\Big\{
27  x_1 - \Big[57 + 6  x_1  (13 + (26 + 51  x_1)  x_2)\Big]  x_3 
\\
&&
+  \Big[72 - 129  x_1 + 3  (68 + x_1  (456 + 25  x_1))  x_2 
\\
&&
+ x_1 (282  +  x_1  (1383 - 470  x_1))  x_2^2\Big]  x_3^2 
\\
&&
+  \Big[105 + 330  x_1 - 714  x_2 - 3  x_1  (200 - 489  x_1)  x_2
\\
&&
-  2  (75 + x_1  (1557 + 912  x_1 + 470  x_1^2))  x_2^2 
\\
&&
- 120  x_1  (3 + 9 x_1  - 16  x_1^2)  x_2^3\Big]  x_3^3 
+  9  \Big[2  x_1^3  x_2^2  (1 - 2  x_2)  (13 + 34  x_2) 
\\
&&
+  3  x_1^2  x_2  (1 - 2  x_2)  (1 - 23 x_2 + 4  x_2^2) 
\\
&&
-  6  x_1  (1 - 2  x_2)  (3 + 36 x_2 + 17  x_2^2 ) 
-  6  (2 - 2 x_2 - 19  x_2^2)\Big]  x_3^4 
\\
&&
+   27  x_2  (1 - 2  x_2)  
\Big[18 + 18 x_1 x_2 - 33 x_1^2 
\\
&&
+  2 x_1^2  x_2  (33 - 9  x_2 +8  x_1  (1 -2  x_2))\Big]  x_3^5
\Big\}
\\
&&
+ 48 \nu x_2^2 x_3^3
\Big\{
18 - 29  x_1
\\
&&
 - \Big[ 21 +  4  (10  x_1 + 12 x_2 
 - 19  (2 - x_1)  x_1  x_2) \Big] x_3 
\\
&&
+  \Big[42 + 106  x_2 +  252 x_1  - 23 x_1^2 x_2 (7 - 10  x_2) 
\\
&&
-  6 x_1 x_2  (3 + 20  x_2)\Big]  x_3^2 
-  \Big[83 +  230  x_2 -  41 x_1
\\
&&
 +  5 x_1  x_2  (134 - 20  x_2 - 3  x_1  (9 + 10  x_2))\Big]  x_3^3 
\\
&&
-  9  \Big[12 - 54  x_2 +  x_1  (1 -  2  x_2) 
(53 - (24 + 41  x_1)  x_2)\Big]  x_3^4 
\\
&&
+   81  (1 - 2  x_2)  \Big[2 + 3  x_1 - x_1  (4 + x_1)  x_2\Big]  x_3^5
\Big\}\,,
\\
&&\\
&&\\
P_{\rm A;\,  box; +}^{(2)} &=&\;\;\;
  - 6 s^3 
 \Big[1 + x_1  (1 - 2  x_2)\Big] x_2^3 x_3^6 x_{123}
\\
&&
\times \Big[8 - 5  x_2 +8  x_1 -x_1  (6 + 5  x_1)  x_2 - 6  x_3 
\\
&&
-  6  x_1  (1 - 2  x_2)  x_3 
+  9  x_2  (1 + x_1  (1 - 2  x_2))^2  x_3^2\Big]
\\
&&
-  6  s \nu^2 (1 - x_1)^2  x_2^3  x_3^6 x_{123}
\\
&&
\times \Big[8 - 6  x_3 + x_2 (1 +  x_1  (1 - 2  x_2))  
(5 - 3  x_3  (4 - 9  x_3))\Big]
\\
&&
 +12  s^2 x_2^2 x_3^4 x_{123}
\Big\{2  (2 + x_2 +  2 x_1 - (6 - x_1) x_1  x_2) 
\\
&&
-  12 \Big[1 + x_1(1 - 2  x_2) \Big] x_3 + \Big[ 13 (2 - x_2) 
\\
&&
+ 2 x_1  (1- x_2)  (13 - 16  x_2) 
- x_1^2 x_2  (13 -  32 x_2 +  32 x_2^2)\Big]  x_3^2 
\\
&&
-  12  \Big[1 + x_1 + x_2 + x_1^2  (1 - 2  x_2)^2  x_2 
- 4  x_1  x_2^2\Big]  x_3^3
\\
&&
 +  27  x_2  \Big[1 + x_1 (1 - 2  x_2)\Big]^2  x_3^4\Big\}
\\
&&
 -  12 \nu^2 (1 - x_1)^2  x_2^3  x_3^4 x_{123}
 \Big\{2 + 12  x_3 - 17  x_3^2 + 24  x_3^3 - 27  x_3^4 \Big\}
\\
&&
+ 24 s  x_2  x_3^4 x_{123} 
\times
\Big\{6  (2 - x_3)  (1 -  x_3)  x_3 
\\
&&
+ 2 x_1 x_2^2 \Big[ 23 - 24 x_3 + x_3^2 (2-3 x_3)(2-9  x_3)\Big] 
\\
&&
+  (1 + x_1)  x_2  
\Big[-23 + 24 x_3 -4  x_3^2 + 3 (8-9 x_3)  x_3^3\Big]\Big\}
\\
&&
+ 144  x_2  (1 - x_3)^2  x_3^4 x_{123}
\Big[ 5 + x_3 (2 + 3  x_3)\Big]\,,
\\
&&\\
&&\\
P_{\rm A;\,  box; -}^{(2)} &=&\;\;\;
6 \nu^3 (1-x_1)^3 x_2^4 x_3^6 x_{123}
\Big[5-6 x_3+9 x_3^2\Big]
\\
&&
+ 6  s^2 \nu (1-x_1) x_2^3 x_3^6 x_{123}
\times \Big\{
16 + 16  x_1 - 5  x_2 - 22  x_1  x_2 - 5  x_1^2  x_2 
\\
&&
-  6  \Big[ 2 + (1 + x_1)  x_2 - 2  x_1  x_2^2\Big]
      \Big[ 1 + x_1  (1 - 2  x_2)\Big]  x_3 
\\
&&
+  27  x_2  \Big[1 + x_1  (1 - 2  x_2)\Big]^2  x_3^2\Big\}
\\
&&
- 24 s \nu (1-x_1) x_2^2 x_3^4 x_{123}
\times \Big\{ 2 - 6 x_3 + (13 - 6 x_3) x_3^2   
\\
&&
- x_2 \Big[ 1 + x_1 (1 - 2 x_2) \Big]  
\Big[6 -2  x_3 + 9  x_3^2  (2 - 3  x_3) \Big]  x_3 \Big\}
\\
&&
+ 72 \nu (1-x_1) (1-x_3) x_2^2 x_3^4 x_{123}
\Big[5 - 7 x_3 + (1-9 x_3) x_3^2\Big]\,.
\end{eqnarray*}

\subsection{The polynomials $P_B$}

\begin{eqnarray*}
P_{{\rm B;\, acn}(1)} &=& \;\;\;
144 x_3 (1-x_3) \Big[x_+ - \frac{\nu}{s} x_- \Big]\,,
\\
&&\\
P_{{\rm B;\,  acn}(2)} &=& \;\;\;
\frac{12 \nu^2}{s} (1-x_3)^2 \Big[(3-5 x_3-x_3^4) x_-^2+6 x_3^2 x_+\Big]
\\
&&
+ \frac{24 \nu x_- }{s} (1-x_3)^2 
( 7 - 12 x_3 + 3 x_3^2 + 4 x_3^3 - 5 x_3^4)
\\
&&
- \frac{48}{s} x_3 (1-x_3) 
( 14 + 19  x_3 - 43  x_3^2 + 29  x_3^3 - 10  x_3^4)
\\
&&
+ 12 s (1-x_3)^2 
\Big\{ x_+   (1 + 3  x_+) - 5  x_3  x_+^2 + 6  x_3^2  x_-^2
\\
&&
- x_3^3  \Big[ 9  x_-^2 - x_+  (2 + 3  x_+) \Big]
+  x_3^4 \Big[ 6  x_-^2 - x_+  (3 + x_+) \Big]  \Big\}
\\
&&
- 12 \nu x_- (1-x_3)^2 
\Big[1 + 6  x_+  - 10  x_3  x_+ +  6  x_3^2  (1 + x_+)
\\
&&
+  2 x_3^3  (1 - 3  x_+)  - x_3^4  (3 - 4  x_+)\Big]
\\
&&
-24  (1 - x_3)^2  \Big[ 7  x_+  -  2  (1 + 6  x_+) x_3
\\
&&
 -  (1 - x_+)  x_3^2  (15 - 26  x_3)  - (15 - 13  x_+)  x_3^4\Big]\,,
\\
&&\\
P_{{\rm B;\, acn}(3)} &=& \;\;\;
\frac{3 (s-\nu)^2}{s}  (1-x_3)^5 (1+x_3)
\\
&&
+ \frac{6 (s-\nu)}{s} (1-x_3)^3 
\Big[ 3 + 5  x_3 + 2  x_3^2 (1 - x_3) \Big]
\\
&&
+ \frac{12}{s}  x_3 (1-x_3) (2-x_3)
\Big[3 (1+x_3) + x_3^2 (1-x_3) \Big]\,,
\\
&&\\
P_{{\rm B;\, acn}(4)} &=& 
- \frac{24 (s-\nu)^2}{s}  x_1^2 (1-x_3)^4 (2+x_3)
\\
&&
- \frac{48 (s-\nu)}{s}  x_1 (1-x_3)^2 \Big[ 2 + x_3 (4-3 x_3) \Big]
\\
&&
- \frac{96}{s} x_3 (1-x_3) (2-x_3) \Big[ 2+(2-x_3) x_3 \Big]\,,
\\
&&\\
P_{{\rm B;\, acn}(5)} &=& \;\;\;P_{{\rm B;  acn}(3)}|_{\rm t\leftrightarrow u}\,,
\\
&&\\
P_{{\rm B;\, acn}(6)} &=&\;\;\;
P_{{\rm B;\, acn}(4)}|_{\rm t\leftrightarrow u,  x_1\leftrightarrow x_2}\,,
\\
&&\\
&&\\
P_{\rm B;\,  ver} &=&\;\;\;
 8 s  x_2^2  (1 - 2  x_2)  x_3^4  
\Big\{ 54 -  8  (69 + 50  x_2)  x_3 
\\
&&
+  15  \Big[ 70 + x_2  (94 + 77  x_2) \Big]  x_3^2
\\
&& 
-  180  \Big[ 3 + x_2  (4 + 23  x_2)\Big]  x_3^3 
  - 405 x_2  (2 - 9  x_2)  x_3^4
\Big\}
\\
&&
- 96  x_2^2  (1 - 2  x_2) x_3^4 
 \Big[25 - 144  x_3 + 500  x_3^2 - 780  x_3^3 + 405  x_3^4\Big]\,,
\\
&&\\
&&\\
P_{\rm B;\, box; +}^{(1)} &=&
-  \frac{12 \nu^2}{s} x_2^3 (1-x_3) x_3^4 
 \Big\{ 20 + 16  x_1  (7 + 9  x_1) 
\\
&&
-  98  x_3  -  x_1  (574 + 573  x_1)  x_3 
\\
&&
+  6 \Big[32 + x_1  (166 + 117  x_1)\Big]  x_3^2 
   - 27  \Big[ 6 + x_1  (18 + 11  x_1)\Big]  x_3^3\Big\}
\\
&&
-  \frac{288}{s} x_2 (1-x_3)^2 x_3^2 
\Big\{ 7 -17 x_3 + 14 x_3^2 - 17 x_3^3 +  27 x_3^4 \Big\}
\\
&&
-12 s  x_2^2  x_3^3  
\Big\{
  -18  x_1 + \Big[38 + 4 (51  x_1 - 5  x_2 + 45  x_1^2  x_2)\Big]  x_3 
\\
&&
-  \Big[156 - 50  x_2 
 + 322  x_1 + 586  x_1  x_2 + 5 x_1^2 x_2 (85 + 74 x_2)\Big]  x_3^2 
\\
&&
+  5  \Big[38 + 18  x_2 
   + 4 x_1 + 278 x_1 x_2 + 76 x_1  x_2^2
\\
&& 
-  3 x_1^2 x_2  (7 -  4  x_2  (17 + 7  x_2))\Big]  x_3^3 
\\
&&
-  3 \Big[24 + 94  x_2 -36 x_1 + 130 x_1 x_2   + 424 x_1  x_2^2 
\\
&&
-  3  x_1^2 x_2  (1 - 2  x_2)  (37 +  76  x_2)\Big]  x_3^4 
\\
&&
+   27  x_2  \Big[6 -  x_1  (1 - 2  x_2)  
   (18 - 11  x_1  (1 - 2  x_2))\Big]  x_3^5\Big\}
\\
&&
+ 48  x_2  x_3^2  
\Big\{ 6 - (34  + 25  x_1  x_2) x_3 
\\
&&
+  \Big[97 - x_2  (7 - 16  x_1  (7 + 4  x_2))\Big]  x_3^2 
\\
&&
-   2  \Big[79 + x_2  (13 + x_1  (191 + 118  x_2))\Big]  x_3^3 
\\
&&
+   5  \Big[25 + x_2  (45 + x_1  (101 + 110  x_2))\Big]  x_3^4 
\\
&&
-  3  \Big[12 +  x_2  (118 - x_1  (7 - 284  x_2))\Big]  x_3^5 
\\
&&
+  81  x_2  \Big[2 - 3 x_1 (1 - 2  x_2)\Big]  x_3^6\Big\}\,,
\\
&&\\
&&\\
P_{\rm B;\, box; -}^{(1)} &=&\;\;\;
\frac{48 \nu}{s} x_2^2 (1-x_3) x_3^3 
\Big\{ 14 + 11  x_1 - 9  (7 + 9  x_1)  x_3 
\\
&&
+  (127 + 373  x_1)  x_3^2 
 - 12  (19 + 46  x_1)  x_3^3 + 81  (2 + 3  x_1)  x_3^4\Big\}
\\
&&
+ 24 \nu  x_2^2  x_3^3  
\Big\{-9  x_1 + \Big[19+2  x_1  (41 + (38 + 81  x_1)  x_2)\Big]  x_3 
\\
&&
- \Big[78 + 34  x_2 +189 x_1 
\\
&&
+  x_1 x_2  (524 + 84  x_2 + 3  x_1 (193 + 59  x_2))\Big]  x_3^2 
\\
&&
+  5  \Big[19 + 38  x_2 + 34 x_1 
+  x_1 x_2  (224 + 78  x_2 + 9  x_1  (9 + 20  x_2))\Big]  x_3^3 
\\
&&
- 3  \Big[2  (6 + 53  x_2) + 18 x_1  
\\
&&
+  x_1  x_2  (240 - 111  x_1 + 248  x_2 + 447  x_1  x_2)\Big]  x_3^4 
\\
&&
+ 27 x_2 \Big[6 + 18 x_1 x_2 - 11 x_1^2 (1 - 2  x_2)\Big] x_3^5
\Big\}\,,
\\
&&\\
&&\\
P_{\rm B;\, box; +}^{(2)} &=&
- \frac{12 \nu^2}{s} (1-x_1)^2 x_2^3 (1-x_3)^2 x_3^4
\\
&&
\times \Big[ 2 + 12  x_3 - 17  x_3^2 + 24  x_3^3 - 27  x_3^4\Big]
\\
&&
+ \frac{144}{s} x_2 (1-x_3)^4 x_3^4 \Big[5+x_3 (2+3 x_3)\Big]
\\
&&
+  12 s  x_2^2  (1 - x_3)^2  x_3^4  
\Big\{ 2  (2 + x_2 +  x_1  (2 - (6 - x_1)  x_2))  
\\
&&        
- 12  \Big[ 1 + x_1  (1 - 2  x_2)\Big]  x_3 
+\Big[13  (2 - x_2) + 26 x_1
\\
&&
- x_1 x_2 (58 - 32 x_2 + x_1 (13 - 32 (1 - x_2) x_2))\Big]  x_3^2 
\\
&&
-  12  \Big[1 + x_1 + x_2 + x_1^2  (1 - 2  x_2)^2  x_2 - 
              4  x_1  x_2^2\Big]  x_3^3 
\\
&&
+ 27  x_2  \Big[1 + x_1  (1 - 2  x_2)\Big]^2  x_3^4
\Big\}
\\
&&
- 6 s^2  x_2^3  
(1 + x_1  (1 - 2  x_2))  (1 - x_3)^2  x_3^6 
\\
&& 
\times \Big\{ 8 -  6  x_3 - x_2  (5 - 9  x_3^2) 
   - x_1^2  x_2  (5 - 9  (1 - 2  x_2)^2  x_3^2) 
\\
&&
+  2  x_1  \Big[ 4 - 3  x_3 - 18  x_2^2  x_3^2 
    -3  x_2  (1 - 2  x_3 - 3  x_3^2)\Big]\Big\}
\\
&&
- 6 \nu^2 (1 - x_1)^2  x_2^3  (1 - x_3)^2  x_3^6  
\\
&&
\times \Big\{ 8 - 6  x_3 
 + x_2  \Big[1 + x_1  (1 - 2  x_2)\Big] 
         \Big[5 - 3  x_3  (4 - 9  x_3)\Big]\Big\}
\\
&&
+ 6  x_2  (1 - x_3)^2  x_3^4  
\Big\{24  (2 - x_3)  (1 - x_3)  x_3 
\\
&&
- 4 (1 + x_1) x_2 
\Big[23 - x_3 (24 - x_3 (2 - 3  x_3) (2 - 9  x_3))\Big]
\\
&&
+ 8 x_1 x_2^2 \Big[23 - x_3 (24 - x_3 (2 - 3 x_3) (2 - 9  x_3))\Big]
\Big\}\,,
\\
&&\\
&&\\
P_{\rm B;\, box; -}^{(2)} &=&\;\;\;
 \frac{6 \nu^3}{s} (1-x_1)^3 x_2^4 (1-x_3)^2 x_3^6 
(5 - 6  x_3 + 9  x_3^2)
\\
&&
+ \frac{72 \nu}{s} (1 - x_1) x_2^2 (1 - x_3)^3  x_3^4  
\Big[ 5 -  x_3  (7 - x_3  (1 - 9  x_3))\Big]
\\
&&
+ 6 s \nu (1 - x_1) x_2^3 (1 - x_3)^2 x_3^6  
\Big\{ 16 (1 + x_1) - 5  x_2 - 22  x_1  x_2 
\\
&&
-  5  x_1^2  x_2
 -  6  \Big[ 2 + (1 + x_1)  x_2 - 2  x_1  x_2^2 \Big] 
         \Big[ 1 + x_1  (1 - 2  x_2)\Big]  x_3 
\\
&&
+  27  x_2  \Big[1 + x_1  (1 - 2  x_2)\Big]^2  x_3^2 \Big\}
\\
&&
- 24 \nu  (1 - x_1)  x_2^2  (1 - x_3)^2  x_3^4  
\Big\{ 2 - x_3  \Big[6 - x_3  (13 - 6  x_3) 
\\
&&
+  (1 + x_1)  x_2  (6 - x_3  (2 - 9  (2 - 3  x_3)  x_3)) 
\\
&&
- 2  x_1  x_2^2  ( 6 -  x_3  (2 - 9  x_3  (2 - 3  x_3)))\Big]
\Big\}\,.
\end{eqnarray*}

\end{document}